\documentclass{article} 
\usepackage{collas2023_conference,times}
\usepackage{easyReview}

\usepackage{amsmath,amsfonts,bm}

\def\eqref#1{equation~\ref{#1}}

\def\1{\bm{1}}

\DeclareMathAlphabet{\mathsfit}{\encodingdefault}{\sfdefault}{m}{sl}
\SetMathAlphabet{\mathsfit}{bold}{\encodingdefault}{\sfdefault}{bx}{n}

\usepackage{hyperref}
\hypersetup{
    colorlinks=true,
    linkcolor=red,
    filecolor=magenta,
    urlcolor=blue,
    citecolor=purple,
    pdftitle={Overleaf Example},
    pdfpagemode=FullScreen,
    }

\title{Autotelic Reinforcement Learning \\in Multi-Agent Environments}

\author{Eleni Nisioti \thanks{Equal contribution}   \quad Elias Masquil \thanks{Work done during internship at the Flowers Team} \footnotemark[1]  \quad Gautier Hamon \footnotemark[1]   \quad \textbf{Clément Moulin-Frier} \\
Flowers Team, Inria and Ensta ParisTech, Bordeaux, France 
}

\preprintcopy 

\usepackage{xspace}

\usepackage{algorithm,algpseudocode}
\newcommand{\algname}{Goal-coordination game\xspace}
\newcommand{\probname}{Dec-IMSAP\xspace}
\newcommand{\probnamelong}{Decentralized Intrinsically Motivated Skills Acquisition Problem\xspace}
\usepackage{subcaption}
\usepackage{comment}

\definecolor{revise}{rgb}{0, 0, 0}

\begin{document}

\maketitle

\begin{abstract}


In the intrinsically motivated skills acquisition problem, the agent is set in an environment without any pre-defined goals and needs to acquire an open-ended repertoire of skills.
To do so the agent needs to be \textit{autotelic} (deriving from the Greek \textit{auto} (self) and \textit{telos} (end
goal)): it needs to generate goals and learn to achieve them following its own intrinsic motivation rather than external supervision.
Autotelic agents have so far been considered in isolation. But many applications of open-ended learning entail groups of agents. 
Multi-agent environments pose an additional challenge for autotelic agents: to discover and master goals that require
cooperation agents must pursue them simultaneously, but they have low chances of doing so if they sample them
independently.
In this work, we propose a new learning paradigm for modeling such settings, the \probnamelong (\probname), and employ it to solve cooperative navigation tasks.
First, we show that agents setting their goals independently fail to master the full diversity of goals.
Then, we show that a sufficient condition for achieving this is to ensure that a group \textit{aligns} its goals,  i.e., the agents
pursue the same cooperative goal.
Our empirical analysis shows that alignment enables specialization, an efficient
strategy for cooperation.
Finally, we introduce the \algname, a fully-decentralized emergent communication algorithm,
where goal alignment emerges from the maximization of individual rewards in multi-goal cooperative environments and show
that it is able to reach equal performance to a centralized training baseline that guarantees aligned goals.
To our knowledge, this is the first contribution addressing the problem of intrinsically motivated multi-agent goal
exploration in a decentralized training paradigm.

\end{abstract}

\section{Introduction}
Many multi-agent scenarios require the cooperation of agents with a rich diversity of skills.
Multi-player games such as StarCraft~\citep{https://doi.org/10.48550/arxiv.1708.04782} and Capture the Flag~\citep{doi:10.1126/science.aau6249} and real-world scenarios such as cooperative navigation in teams of robots~\citep{wangModelbasedReinforcementLearning2020}, require agents that can
coordinate their actions in the face of continuously-arising new challenges.
When alone, a reinforcement learning (RL) agent can acquire a wide diversity of skills by being \textit{goal-conditioned}~\citep{liu2022GoalConditionedReinforcementLearning} and \textit{intrinsically motivated}~\citep{oudeyer2009intrinsic,NIPS2004_4be5a36c,colas2022autotelic}:
the former means that the agent can pursue different goals at different times and conditions its learning on its
current goal, while the latter means that these goals are generated by the agent using some internal reward mechanism instead of being externally set by the human designer.
Such agents have been termed autotelic~\citep{colas2022autotelic}.
But what happens when you place multiple autotelic agents in the same room, expecting them to
autonomously discover all the room's affordances?
We argue that you will stumble upon a challenge: for agents independently generating their own goals, the
probability of sampling the same one reduces dramatically with the size of the goal space.
Thus, we expect that these agents will fail to master goals that require cooperation (such as lifting a heavy box),
as they will collectively pursue them rarely and receive a noisy training signal due to the fact that the goals of others are not directly observable.
In this work, we introduce a new type of problem for multi-agent RL, the \probnamelong (\probname), to capture such
settings, propose a decentralized algorithm for tackling it, the \algname, and evaluate it in a cooperative
navigation task~\footnote{We provide code for reproducing the experiments presented in the paper at 
\url{https://github.com/Reytuag/imgc-marl}}.

\begin{figure}
    \centering
    \includegraphics[width=0.6\textwidth]{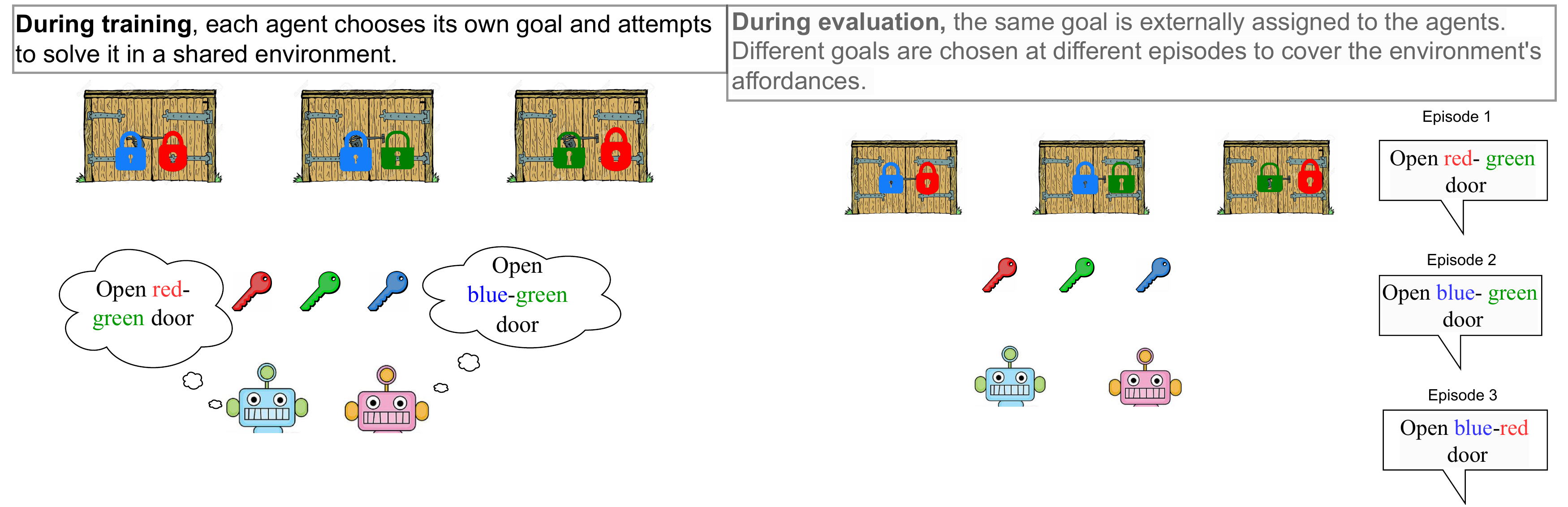}
\caption{Illustrative example of learning in a \probname: \textmd{two agents are in a shared environment where goals have the form of doors that open upon matching each lock with the key of the same color. An agent can carry at most one key, so it takes two to open a door. (Left) As agents are sampling their own goals, the group may experience episodes that cannot be solved by at least one agent: if the blue agent picks the red key and the pink agent picks the green then the blue agent will succeed and the pink will fail. (Right) During evaluation agents are assigned with the same goal.}
}
\label{fig:intro_fig}
\end{figure}

For single-agent settings, autonomous skill discovery has been formalized as the Intrinsically Motivated Skills Acquisition Problem~\citep{colas2022autotelic}.
To meaningfully extend it to multi-agent settings we need to consider environments that require
cooperation: some of the goals will be cooperative, i.e., at least one other agent needs to act for the agent to achieve its  goal, while the rest will be independent, i.e., they can be achieved independently of others.
To study the \probname we propose a new training/evaluation paradigm: during the training phase agents are autonomously
setting their own goals and learning to achieve them in a fully-decentralized manner, while, during evaluation,
we externally provide agents with the same cooperative or individual goal, ensuring that a wide diversity of goals is
tested across evaluation episodes.
We provide an illustrative example of this problem setting in Figure~\ref{fig:intro_fig}.

Intrinsic motivation originated in the field of cognitive science, with studies focusing on human infants due to their impressive ability to efficiently learn new skills~\citep{oudeyer2009intrinsic}.
Explanations rely on exploratory play, during which infants generate their own goals and learn how to achieve them for the mere purpose of discovering new learning situations~\citep{berlyne1966curiosity,gopnik1999scientist}.
While studies primarily consider a single human subject, some study infants engaging in cooperative play and show that they can 
plan alongside others
~\citep{warneken2014young,rubin1978free,hamann2012children}.
According to theories of human social intelligence~\citep{tomasello2007shared}, we may owe our ability to cooperate more extensively than other species to our \textit{shared intentionality}: to solve tasks that require cooperation we attend to the same goal with others and know that we are doing so.



Does shared intentionality play an equally important role in groups of artificial agents and, if so, how can we guarantee
it in a fully-decentralized training regime?
This is the main research question we aim to address with our study of the \probname.
{\color{revise} As we more concretely explain in Section \ref{sec:formal}, we are motivated by real-world applications, such as groups of cleaning robots or disaster robotics, where a group needs to adapt to a diversity of tasks, some of which may require cooperation.} To address this question empirically, we {\color{revise} study a simplified two-player setting with goal-conditioned RL agents~\citep{colas2022autotelic} that sample goals randomly from a fixed, pre-defined set. Such agents have been previously extended to multi-agent settings assuming external supervision during training~\citep{yang2018cm3}, thus not considering autonomous learning.} 
We measure the degree of shared intentionality as \textit{goal alignment}, a metric quantifying the percentage of training
episodes during which two agents pursue the same cooperative goal.
First, we artificially control for the level of alignment and observe it is highly correlated with performance.
Then, we propose the \algname, a fully-decentralized emergent communication algorithm that enables goal coordination during training.
Under the \algname, before acting, an agent, chosen at random, takes the role of a leader, samples its own goal and communicates a message to the follower. which selects its own goal based on it.
Crucially, the agents learn how to map goals to messages and vice versa by purely maximizing their individual rewards.
{\color{revise} By coordinating in the message, rather than the goal space, agents using the \algname can align their goals even if they employ different goal representations.}
We show that alignment emerges so that the population reaches equal performance to a centralized
setting that guarantees alignment.
To get a clearer understanding of the temporal dynamics of the \algname, we
analyze the co-evolution of messages, goal alignment and group rewards and discover that interesting collective behaviors emerge. Our contributions are:
\begin{enumerate}
    \item the formulation of the \probname, a new type of problem for studying intrinsic motivation in multi-agent systems with goal-conditioned RL agents;
    \item a detailed analysis on the impact of goal alignment between agents in the \probname;
    \item an algorithm for solving the \probname, the \algname, that enables agents in a group to acquire a large repertoire of cooperative skills in a fully-decentralized setting by learning how to communicate about their respective goals.
\end{enumerate}

We discuss related works in Section \ref{sec:related} and, in Section \ref{sec:background}, provide definitions from existing works in single-agent and multi-agent RL that we built upon to formulate the \probname. We, then, present our formal definition of the \probname and the algorithm that we propose for solving it, the \algname in Section \ref{sec:framework}. In Section \ref{sec:results}, we empirically analyze the behavior of groups of agents, where we employ cooperative navigation tasks as instances of the \probname. Finally, we discuss limitations and future directions for our work in Section \ref{sec:discuss}. 


\section{Related Works}\label{sec:related}
{\color{revise} Early in RL development, real-world applications, such as human-interfacing robots, pushed for algorithms that can solve, not just a static task, as classically assumed by the framework of Markov Decision Processes, but tasks that change with time~\citep{kaelbling1993LearningAchieveGoalsa}.
This problem setting, termed multi-goal RL, has been formulated under frameworks such as options and skills~\citep{sutton1999MDPsSemiMDPsFramework,konidaris2009SkillDiscoveryContinuous}, goal-conditioned policies~\citep{andrychowicz2017hindsight,colas2022autotelic} and universal value function approximators~\citep{schaul2015UniversalValueFunction}.
Multi-goal RL has been extended to multi-agent settings, such as cooperative navigation in fleets of robots, under the framework of multi-goal Markov games~\citep{yang2018cm3,mordatch2018emergence}: a group of agents employing goal-conditioned policies is trained under the supervision of a human designer that selects which goal each agent needs to pursue during each training episode.

Real-world applications soon posed another, more stringent requirement on RL: tasks do not just vary with time but may change in unpredictable ways. 
Thus, autonomously discovering new tasks and adapting online to them became part of the problem~\citep{nguyen2014SociallyGuidedIntrinsic,parisi2019ContinualLifelongLearning}.
Inspiration for tackling this setting was found in child development, in particular in exploratory play, during which infants showcase an impressive ability to self-generate their own goals and autonomously learn how to achieve them.
This mechanism has been termed intrinsic motivation and has inspired a variety of unsupervised learning objectives for RL algorithms in both single-agent~\citep{https://doi.org/10.48550/arxiv.1810.06284,pathak2017CuriositydrivenExplorationSelfsupervised,achiam2017SurpriseBasedIntrinsicMotivation,schmidhuber2009DrivenCompressionProgress} and multi-agent~\citep{jaques2019social} settings.
Not all intrinsically-motivated agents are goal-conditioned~\citep{pathak2017CuriositydrivenExplorationSelfsupervised,schmidhuber2009DrivenCompressionProgress,jaques2019social}.
The ones that are, termed autotelic~\citep{colas2022autotelic}, are particularly interesting for real-world applications, as they precisely capture multi-goal settings with goals set and mastered autonomously by the agent.
If we attempt to transfer autotelic RL to multi-agent settings, we see that multi-goal Markov games are not adequate.
By assuming that an external supervisor is deciding which goals will be pursued and how they will be divided among agents, it bypasses the main question in open-ended settings: how can the agents set goals autonomously?
Taking into consideration that some of these goals may require the coordination of multiple agents, autotelic learning becomes more challenging when transferred to multi-agent setups.

Although previous works studied the interaction between autotelic agents and social partners~\citep{nguyen2014socially,colas2020language}, we first propose here to study the coordination of goals set by a group of autotelic agents without any prior knowledge.
For this, we can draw inspiration from the problem of intra-episode action coordination, a long-standing subject in MARL.
For example, in a cooperative navigation task, others may act as obstacles that affect the observations perceived by an agent in an unpredictable way.
To tackle this problem, algorithms may choose to learn a centralized critic while keeping policies decentralized~\citep{foerster2017CounterfactualMultiAgentPolicy,loweMultiAgentActorCriticMixed2020}, model the behaviors of others~\citep{wangModelbasedReinforcementLearning2020} or learn to communicate~\citep{foersterLearningCommunicateDeep2016,lazaridouEmergentMultiAgentCommunication2020,Steels2015}.
This last approach is particularly promising when the group is heterogeneous, as coordination takes space in an abstract, learned space.
Recent emergent communication algorithms have focused on action selection during the episode, while our work considers communication for goal selection before the episode starts.
This is a novel problem that we study in a simplified setting and can benefit from earlier algorithms in the field developed precisely for studying the emergence of shared lexicons in grounded settings~\citep{Steels2015}.
} 


\section{Background}\label{sec:background}

We first describe the problem of autonomous skill acquisition in single-agent settings
as an evolution from classical RL to goal-conditioned and intrinsically-motivated agents in Section~\ref{sec:IMGEP}.
Then, in Section~\ref{sec:GCMARL}, we discuss multi-goal Markov games as a generalization of MARL to goal-conditioned settings with externally-provided goals .

\subsection{Intrinsically motivated goal-conditioned reinforcement learning}\label{sec:IMGEP}
In RL an agent observes an environment and performs actions on it that incur rewards, aiming at
maximizing the rewards it accumulates.
This interaction is commonly formalized as a Markov Decision Processes (MDP): at
each time step $t$ of an episode that lasts for $T$ time steps the agent observes the environmental state $s_t$,
performs action $a_t$ and receives reward $r_t$.
The policy $\pi(a_t|s_t)$, which describes the agent's behavior as a mapping from states to actions, is interactively
learned from experience by maximizing the cumulative reward $G_t= \sum_{t=0}^T \gamma^t r_t$, where $\gamma$ is a
parameter quantifying how heavily future rewards are discounted~\citep{sutton2018reinforcement}.
Formally, we denote an MDP as a tuple $(\mathcal{S}, \mathcal{A}, \mathcal{T}, \rho_0, R)$, where the state space $\mathcal{S}$
and action space $\mathcal{A}$ indicate all possible configurations for the state and action respectively, $\mathcal{T}(s_{t+1}|s_{t},a_{t})$
is the transition function that controls the distribution of the next state $s_{t+1}$ from the current state $s_t$
when the agent takes action $a_t$, $\rho_0$ is the distribution over the initial states, and the reward
function $R(s_t, a_t)$ determines the reward that an agent receives at each time step for a given state-action
combination.

Agents may need to reward themselves differently based on the task they are currently occupied with.
For example, if we imagine a cleaning robot in a household, then the action ``turn on the oven''
should be rewarded if the robot's task is to prepare food but penalized if the task is to make sure the tenant can
safely leave for a weekend trip.
To expand MDPs to suit this multi-task nature of problems, the goal-conditioned RL paradigm introduces the
notion of a \textit{goal} and conditions the reward function and policy on it.
Formally, a goal $g$ is a tuple $(z^g, R^g)$, where $z^g$ is a goal embedding, $R^g$ denotes the goal-conditioned
reward function and $\pi^g$ the goal-conditioned policy~{\color{revise} \citep{colas2022autotelic,schaul2015UniversalValueFunction,sutton2011HordeScalableRealtime,kaelbling1993LearningAchieveGoalsa}}.
We can, thus, define a multi-goal MDP as a set of MDPs that share $\{ \mathcal{S}, \mathcal{A},  \mathcal{T}, \rho_0 \}$
and differ only in the reward function $R^g$.
We denote the space of possible goals as $\mathcal{G}$.

\begin{figure}
\centering
\includegraphics[width=0.8\textwidth]{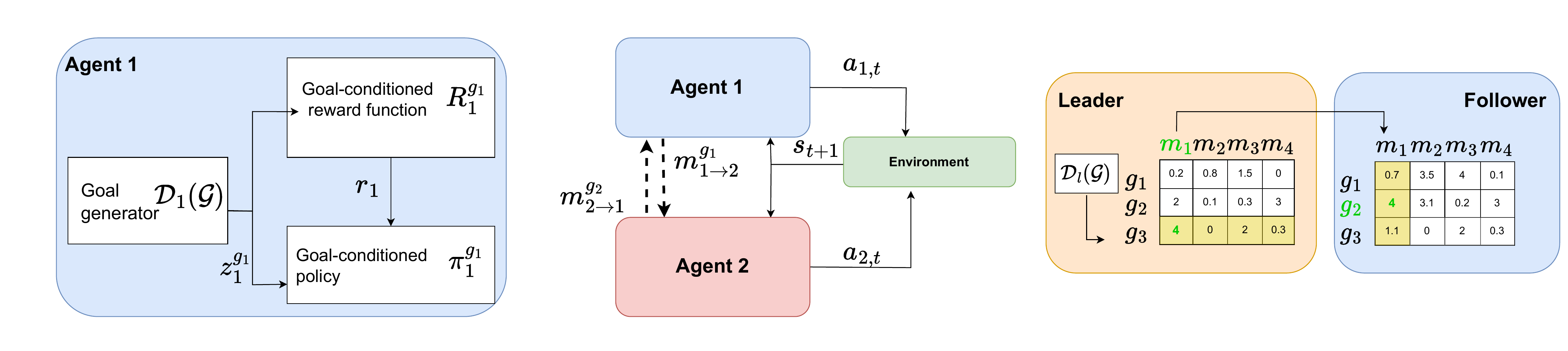}

    \caption{ (Left) Illustration of an autotelic agent equipped with: a goal-sampling distribution $\mathcal{D}_n({\mathcal{G}})$ for selecting its own goal at the beginning of each training episode, the goal-conditioned policy $\pi^g_n$  and reward function $R_{n}^{\mathcal{G}}$. (Middle) Two autotelic agents in a shared environment, trained in a fully-decentralized manner and able to exchange messages $m$ through a discrete communication channel that helps them coordinate their goal selection.
(Right) Illustration of the \algname: \textmd{each agent maintains its own matrix associating goals to messages (see Section \ref{sec:game} for a description of how this matrix is learned). After the leader and follower roles are randomly assigned to the two agents,  the leader samples its own goal (e.g. $g_3$) and transmits the message sampled from a softmax on the corresponding row ($m_1$). The follower samples the goal from a softmax on the corresponding column ($g_2$)}}
    \label{fig:framework}
\end{figure}

Multi-task learning is necessary but not sufficient for open-ended
learning.
In the latter, the agent needs to master not just multiple tasks but a continuously increasing set of
tasks, potentially unknown at the time of the agent's design.
Thus, in contrast to the classical goal-conditioned setting where goals are \textit{externally provided}, the
agent will need to generate them itself, through what is called intrinsically-motivated goal exploration~{\color{revise} \citep{https://doi.org/10.48550/arxiv.1810.06284,pong2020SkewFitStateCoveringSelfSupervised,blaes2020ControlWhatYou,nair2018VisualReinforcementLearning}}.
This setting has been termed as the Intrinsically-Motivated Skills-Acquisition
Problem and the agents that can solve it as autotelic agents \citep{colas2022autotelic}.
Contrary to the classical RL paradigm where a reward function is part of the environment, an autotelic agent
encapsulates the reward function $R^g$, alongside with a mechanism for sampling goals from the goal space, the goal-sampling function
$\mathcal{D}_{\mathcal{G}}$.
We provide an illustration of an autotelic agent on the left of Figure~\ref{fig:framework}.

\subsection{Goal-conditioned multi-agent reinforcement learning}\label{sec:GCMARL}

In multi-agent RL $N$ agents interact in a shared environment, a setting that can be formalized as a Markov Game~\citep{littman1994MarkovGamesFramework}.
The group's behavior is captured by the joint action $\vec{a}_t = \langle a_{1,t}, \cdots, a_{N,t}\rangle$
where $n$ indicates agent's index.
After all actions are executed, the environment returns the next state $s_{t+1}$ and a local reward for each agent $r_{n,t}= R_n(s_t, \vec{a_t})$. To model decentralized learning in a Markov Game we can employ the framework of decentralized partially-observable MDPs
(Dec-POMDPs)~\citep{bernstein2013ComplexityDecentralizedControl}.
\textit{Decentralization} characterizes multi-agent systems where agents do not have access to the observations of others and \textit{partial observability} refers to the fact that this local information may
not be sufficient to infer the environment's state, which now includes the other agents.
To capture partial observability POMDPs introduce the notion of an observation $O_n$ which maps the environmental
state to a local observation for agent $n$.
Formally, a Dec-POMDP is modeled as a tuple ($\mathcal{N}, \mathcal{S}, \{ \mathcal{A}_n \}, \mathcal{T}, \{\mathcal{R}_n\}, \{\mathcal{O}_n \}$),
where $\mathcal{N}$ denotes the set of agents and $\mathcal{A}_n$ and $\mathcal{O}_n$ are the action and observation space of a single agent.

Multi-goal Markov Games extend Markov Games to goal-conditioned settings~\citep{yang2018cm3}.
They arise when we replace the reward function with one conditioned on goals that is shared by all agents: $r_{n,t}= R(s_t, \vec{a}_t, g_n)$.
In multi-goal Markov Games goals are externally provided by a supervisor. Each agent has one fixed goal, only known to itself, and rewards are individual even though the reward function is shared, as they are conditioned on goals.

\section{Autotelic agents in goal-conditioned games}\label{sec:framework}

\subsection{Motivation}
How can a learning framework model a group of agents whose objective is to learn a diversity of goals in a shared
environment without external supervision?
Autotelic learning well captures autonomous skill acquisition but does not consider interactions between multiple
agents.
Multi-goal Markov Games, on the other hand, model interactions of co-existing goal-conditioned agents but do not account
for the fact that agents may be generating their own goals.
We refer to this problem setting as the \textit{\probnamelong} (\probname) and, in the following, provide a formal definition for it.

\subsection{Formalization}\label{sec:formal}
A \probname is modeled as a tuple  $(\mathcal{N}, \mathcal{S}, \{\mathcal{O}_n\}, \{\mathcal{A}_n\}, \mathcal{T}, \{R^g_n\} $, $\{\mathcal{D}_n(\mathcal{G})\})$,
where  $\mathcal{N}$ is the set of $N$ agents, $\mathcal{S}$ is the state space, denoting all the possible configurations of
all $N$ agents and the environment, $O^n$ and $\mathcal{A}^n$ are the observation and action space for a single agent, $\mathcal{T}(s'|s,\vec{a})$ the transition function,
$R_g^n$ is the goal-conditioned reward function and $\mathcal{D}_n(\mathcal{G})$ is the goal-sampling distribution of agent $n$. 
Note that, differently from multi-goal Markov Games, described in Section \ref{sec:GCMARL}, the reward function is not shared among agents. This is because, as we described in \ref{sec:IMGEP}, in intrinsically-motivated learning the reward function is internal to the agent and, thus, may differ across agents.
Also, we assume that the goal space $\mathcal{G}$ which contains all possible goals $g$, is known and identical for all agents.
We illustrate two autotelic agents in a shared environment in the middle of Figure \ref{fig:framework}.

At the beginning of a training episode, each agent $n$
samples its own goal $g_n \in \mathcal{D}_n(\mathcal{G})$, executes its goal-conditioned policy $\pi_n^{g_n}$ and adjusts its behavior to maximize the cumulative reward using goal-conditioned RL.
After a fixed number of training iterations, agents will be evaluated over all possible tasks in a coordinated fashion. By "coordinated" we mean that, during evaluation, agents are assigned with the same, randomly-sampled goal. By doing so, we ensure that there is a fair evaluation of the group's ability to solve all possible cooperative tasks. Agents will be evaluated on the cumulative reward they get and the time they take to solve the goal. 

{\color{revise} 
The \probname is a problem formulation that can well capture the need for autonomous skill acquisition in teams of robots employed in real-world applications. For example, picture a group of assistance robots employed by a company to clean their offices.
Naturally, the team is expected to execute various tasks, some of which may require a single robot while others may require multiple of them (for example carrying a heavy table to another room). How can the company be certain that the robots can execute any possible task when asked to? Following an externally-supervised training paradigm, the company could list all anticipated tasks, assigning a sub-task to each agent~\citep{yang2018cm3}. But this approach quickly becomes impractical once one acknowledges that the list may be large and change in unanticipated ways. 
Under the \probname, we propose an unsupervised training paradigm to exactly tackle these challenges. In this example, the group of agents is left for some time in the offices to discover all their affordances and learn how to solve them. In our proposed solution, the robots can come from different manufacturers, as they learn a communication protocol that allows them to coordinate even if their goal representations differ. 
} 

To solve the \probname the agents must learn how to solve a wide diversity of cooperative goals during training.
Since both goal selection and training are decentralized this is not guaranteed: if agents sample their goals independently then some cooperative goals may not be pursued enough times during training for the group to learn how to achieve them.
In addition, the reward feedback is noisy: even if agents have learned optimal policies for all cooperative goals, they can obtain zero reward if their sampled goals are inconsistent (a case we have illustrated on the left of Figure~\ref{fig:intro_fig}).

\subsection{The \algname}\label{sec:game}
We would like to introduce a process that allows the agents to coordinate their goals without introducing centralization nor pre-existing knowledge within the group and is flexible enough to deal with any behavior arising during training.
To achieve this, we propose an algorithm inspired from the Naming Game, an algorithm originally introduced to help a population of agents invent a shared lexicon~\citep{Steels2015}. 
Our proposed algorithm, whose pseudocode we present in Algorithm \ref{alg:align}, takes place right before an episode starts.
As is common in emergent communication literature, it employs two agents and can be extended by considering a population of
agents and randomly sampling a pair of them at each episode.
Each agent is equipped with a communication matrix $C_n:  |\mathcal{G}|\times|\mathcal{M}| \rightarrow \mathcal{R}$, where $\mathcal{G}$ is the
goal space and $\mathcal{M}$ is a message space, where we consider that both spaces are discrete (we discuss in Section~\ref{sec:discuss}
an extension to continuous spaces).
Each row of matrix $C_n$ corresponds to a different goal $g$ of the agent and each column to a different message $m$.
All values of the tables are initialized with zeros (line 3).
Communication is asymmetric: at the beginning of the goal-coordination round one
agent is randomly chosen to be the leader and the other the follower (lines 8 and 9),
therefore ensuring that each agent takes both roles across episodes.
In what follows we employ underscore $l$ to denote properties of the leader
and underscore $f$ for the follower.
When an agent is the leader, the entries of its matrix answer the question: ``What reward do I expect in this episode if I transmit message $m$ when I have goal $g$? ''.
When an agent is the follower, the question is: ``What reward do I expect in this episode if I choose goal $g$ when I receive message $m$? ''.
Thus, an agent maintains a single matrix that it employs both as a leader (to infer a message given a goal)  and a follower (to infer a goal given a message).

\begin{algorithm}[t]
\scriptsize
\caption{\algname}
\label{alg:align}
\begin{algorithmic}[1]
\State \textbf{Input:} Population: $\mathcal{P} = [n_1, \cdots, N]$, matrix update rate $\alpha$, message space size $M$, goal space size $G$, batch size $B$

\For{agent $n \in \mathcal{N}$} \Comment{Initialize matrices}
\State Initialize $C_n=zeros(G, M)$
\EndFor

\While{not converged}

\State rollouts = []

\For{episode $\in \{1, \cdots, B\}$} \Comment{Collect a batch of episodes}

\State $l = \text{sample}(\mathcal{P})$ \Comment{Randomly select leader}

\State $f = \text{sample}(\mathcal{P}-l)$

\State $g_l= \text{l.chooseGoal()}$ \Comment{Sample goal with $\mathcal{D}_{\mathcal{G}}$}

\State $m_l=\text{softmax}(C_l[g_l, :])$

\State $m_f = m_l$

\State $g_f=\text{softmax}(C_l[:, m_f])$

\State rollouts.append(collectRollout($l,f, g_l, g_f$)) \Comment{Run a single episode}

\EndFor

\For{agent $n \in \mathcal{N}$}
\State Initialize $update_n=zeros(G, M)$;Initialize $norm_n=zeros(G, M)$
\EndFor

\For{rollout $\in$ rollouts}

    \For{agent $n \in \mathcal{N}$}
    \State $n.trainPolicy(rollouts)$ \Comment{Update policies with new experience}
    \EndFor
    \State $n_1, g_1, m_1, r_1 = rollout.l, rollout.g_l, rollout.m_l$ \Comment{Unpack information}

    \State $n_2, g_2, m_2, r_2 = rollout.f, rollout.g_f, rollout.m_f$

    \State $update_{n_1}[g_1, m_1] += r_1$; $norm_{n_1}[g_1, m_1] += 1$;$update_{n_2}[g_2, m_2] += r_2$;$norm_{n_2}[g_2, m_2] += 1$
\EndFor

\For{agent $n \in \mathcal{N}$}
\State $C_n = (1-\alpha)\cdot C_n + \alpha \cdot update_{n}/norm_{n}$ \Comment{Apply matrix update}
\EndFor

\EndWhile

\end{algorithmic}
\end{algorithm}

The leader first samples a goal $g_l$ according to its own goal sampling strategy, $\mathcal{D}_{l}(\mathcal{G})$ (line 10), and, then, transmits the message $m_{l\rightarrow f}$ chosen using a softmax over the corresponding row of $C_l$ (line 11).
The follower receives message $m_{l \rightarrow f}$ and applies a softmax on the corresponding column  $C_f$ to pick its own goal (line 13).
After playing several episodes every agent updates its matrices to reflect the average reward for that specific goal/message association computed on the batch of collected episodes collected (lines 20-29).
Note that during an episode the leader and follower may be pursuing different goals, so although they experience the same episode their rewards may differ.
To ensure that the matrix updates are not too quick for an agent to adapt its policy, we employ an exponential moving average update function with update rate $\alpha$ (line 32). We illustrate a single round on the right of Figure~\ref{fig:framework}.
We should emphasize that agents in the \algname are maximising their individual rewards, conditioned on goals
that may differ and without access to the observation, goal, action and reward of others. The agents may successfully communicate, in the sense that they coordinate their goals, but  will not be rewarded if their policies cannot achieve them.
Vice versa, they may succeed in an episode even if they don’t communicate meaningfully.

\section{Empirical results}\label{sec:results}
\subsection{Setup}
\begin{figure}
    \centering
    \includegraphics[width=0.47\textwidth]{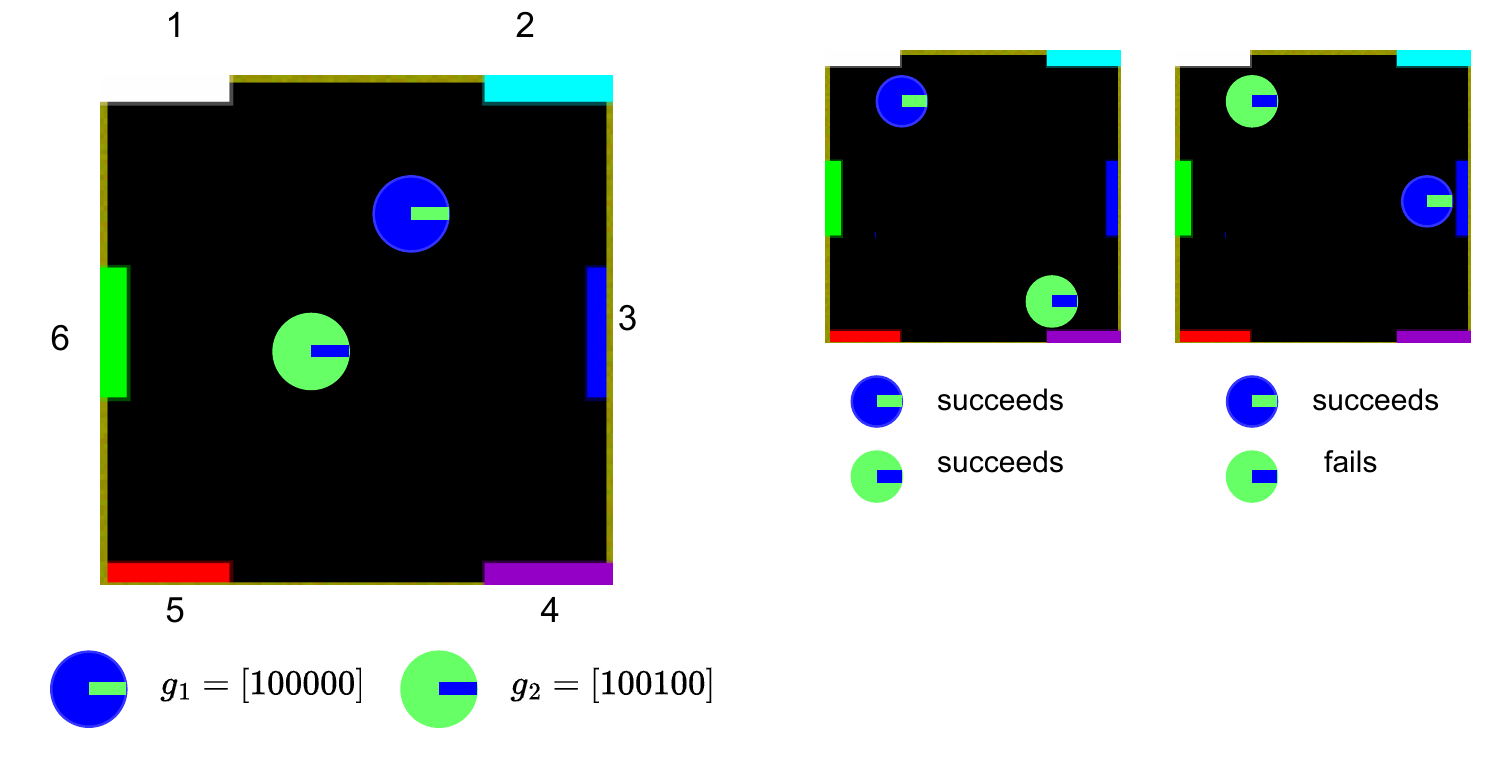}
    \caption{The Cooperative landmarks environment consists of a room with two agents and six landmarks, indicated as colored rectangles. Tasks are formulated as landmarks the agents need to navigate to. During training, the agents sample their own goals that can be individual or cooperative, as the ones chosen by the blue and green agents respectively. An episode may succeed for one agent and fail for the other, the outcome depending on the actions of both.}
    \label{fig:env}
\end{figure}

We study the \probname in the Cooperative landmarks environment that we implemented using Simple Playgrounds
~\citep{Simple-Playgrounds}.
This 2-D environment, illustrated in Figure~\ref{fig:env}, consists of a room with $L=6$ landmarks on its walls and two agents that receive continuous-valued observations about the distance and angle to all landmarks and other agents. They can move by performing discrete-valued actions that control their angular velocity and longitudinal force.
We consider navigation tasks where agents need to reach different landmarks and define goals as vectors of dimension $L$ that are either one-hot or two-hot, the former corresponding to individual goals and the latter to cooperative.
Formally, $g=[x_1, \cdots, x_l], \sum_l[x_l] \in [1,2]$ where $x_l=1$ indicates that landmark $l$ needs to be reached by at least one agent for the goal to be achieved and landmarks are indexed starting from the white one and continuing clockwise (we provide the complete list of goal encodings in Appendix~\ref{app:playground}, alongside a formal definition of the observation and action space).
An episode finishes for an agent once it receives a reward or a time limit is reached. Then, it waits for others to also complete their episode before a new one starts. 
We illustrate an example in Figure \ref{fig:env}: if the blue agent samples goal $[100000]$ and the green $[100100]$, then: a) if blue navigates to the white landmark and green navigates to the purple landmark the episode succeeds for both agents b) if blue navigates to the blue landmark and green to the white landmark then the episode succeeds for the blue agent.
Each agent learns a goal-conditioned policy using PPO with a feedforward policy and uniform goal-sampling distribution $\mathcal{D}({\mathcal{G}})$ (We provide the values of all agent hyper-parameters in Appendix~\ref{app:hypers}).
{\color{revise} To investigate the effect of using more complex intrinsic motivation mechanisms, in Appendix \ref{app:LP} we replace uniform sampling with learning progress~\citep{https://doi.org/10.48550/arxiv.1810.06284}.
We have also studied a baseline that uses a recurrent policy in Appendix \ref{app:recurrent} to investigate whether memory can help agents coordinate.
}
We introduce hyper-parameter $\beta$ for controlling the relative importance between independent and cooperative goals by dividing rewards for individual goals by $\beta$. We do so to model the benefits of cooperation: outcomes that require cooperation often bring larger rewards than outcomes easily solved by a single agent (for example catching a big animal is more rewarding than catching a small one~\citep{skyrms2001StagHunt}).
We set $\beta=2$ in the main paper and study the effect of this hyper-parameter in Appendix~\ref{app:multiplier}. 
To study the effect of environmental complexity, we performed experiments with a smaller environment ($L=3$) in Appendix~\ref{app:3_landmarks}.

In Section~\ref{sec:role_align} we evaluate the role of goal alignment by designing baseline goal-sampling strategies for different levels of it. For a given  $x\%$ desired level of alignment, each agent samples its own goal using $\mathcal{D}({\mathcal{G}})$, but in $x\%$ of the trials we interfere in the sampling procedure and externally provide the agents with the same goal. Therefore, $0\%$ alignment corresponds to autotelic agents sampling their own goals independently of the other at each episode and $100\%$ alignment corresponds to a centralized goal-selection mechanism where a a goal is first sampled externally at the start of each episode and then provided to both agents (similar to the method used by \cite{yang2018cm3}).
We evaluate $0\%$-aligned (also referred to as independent), $50\%$-aligned  and $100\%$-aligned (also referred to as centralized).
{\color{revise} We also evaluate a common method in MARL that follows the centralized training with decentralized execution paradigm (CTDE)~\citep{foerster2017CounterfactualMultiAgentPolicy}, where every critic has access to all goals, actions and observations of the group.
We refer readers to Appendix \ref{app:baselines} for an illustration of how these methods differ in terms of the information available to each agent.
}
Finally, in Section~\ref{sec:game_results} we evaluate the ability of our proposed algorithm, the \algname to reach the performance of the centralized baseline and provide insights into how alignment and performance co-evolve.




\begin{figure}
\centering
\begin{minipage}{0.84\textwidth}
        \centering
    \includegraphics[width=0.8\textwidth]{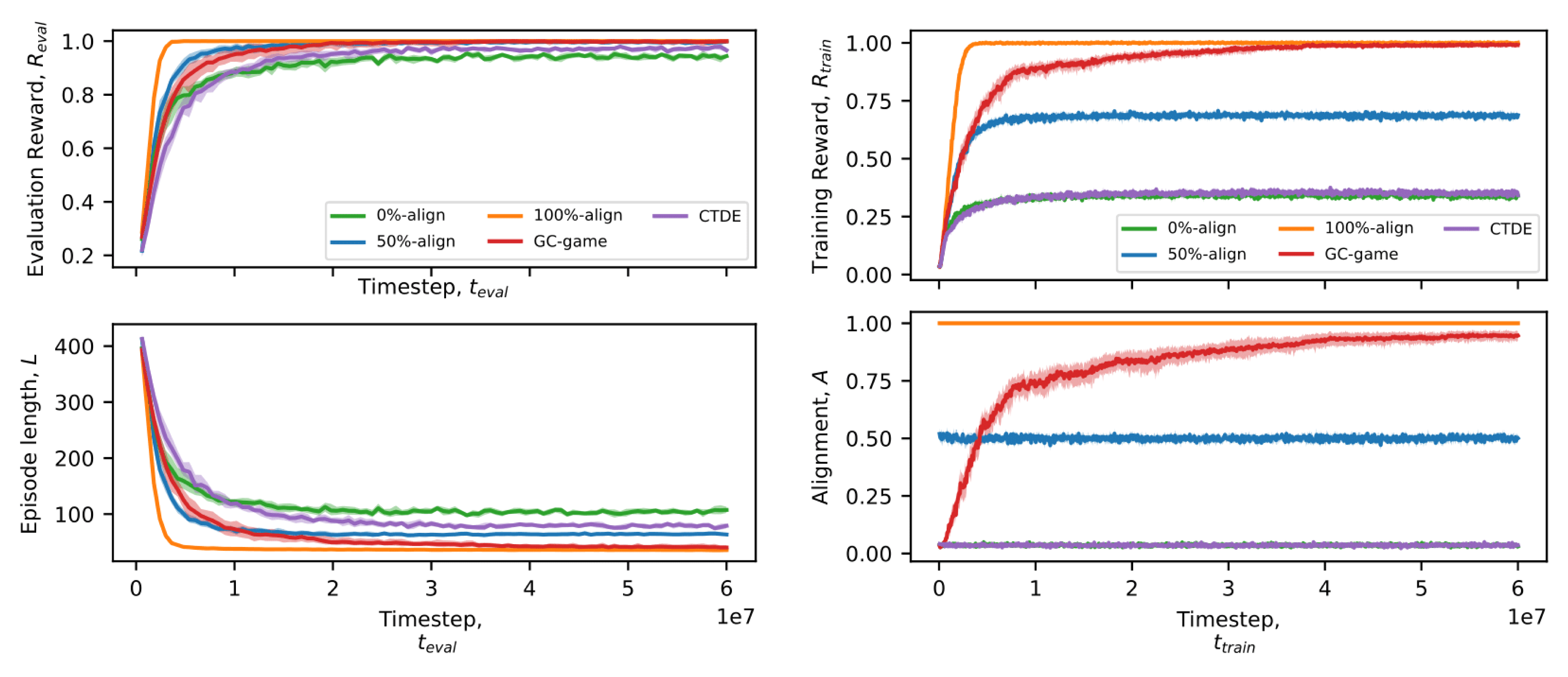}
\end{minipage} \quad
    \caption{Comparison of  baselines with different levels of alignment and the \algname in terms of performance during evaluation (Left) and during training (Right). We present IQM values with stratified bootstrap confidence intervals computed over 20 seeds.}
    \label{fig:performance}
\end{figure}

\subsection{The role of alignment}\label{sec:role_align}

We have hypothesized that agents not aligning their goals during training will not master cooperative goals, as they will collectively pursue them rarely and receive a noisy training signal as the goals of others are not directly observable.
We now examine this hypothesis by comparing the performance during evaluation and training trials between groups of centralized, independent and $50\%$-aligned agents in Figure~\ref{fig:performance}.
In addition to the collected rewards we monitor alignment during training trials and the length of the episode during evaluation trials, where shorter episodes indicate that the group solved the tasks quicker.
We observe that ensuring alignment during training improves performance during evaluation.
In particular, the evaluation reward at the end of training (at time step $426\cdot10^6$) is $\textstyle 0.8277 \pm 0.0436$ for independent, $\textstyle 0.9133 \pm 0.0027$ for $50\%$-aligned and $\textstyle 0.9166$ for centralized. 
Similarly for the episode length, independent requires significantly more time than other methods.
{\color{revise} Our study of the smaller environment in Appendix~\ref{app:3_landmarks} showed qualitatively similar behaviors, with differences between methods being more pronounced. Thus, alignment acquires more significance as the environment becomes more complex.}
The differences in performances are primarily due to cooperative goals; lowering alignment does not have a big impact on the individual goals (we confirm this in Appendix~\ref{app:coop_only}, where we train and evaluate only with cooperative goals and observe similar conclusions as in the current setup but with more visible gap between methods).


As we discussed in Section \ref{sec:formal}, independent agents may fail because: a) as they cannot observe the goals of others and may choose incompatible goals during a training episode, the reward signal does not allow to discriminate between an infeasible episode and a feasible episode where the agents acted sub-optimally b) a large part of the training episodes is infeasible so the agents require more training time compared to centralized. To find out which of the two is the case we evaluated an additional method that we refer to as "both-goals": agents sample their goals independently but we provide both of them to each agent. In this way, the agent can learn which combinations of goals are incompatible and, thus, denoise the training signal.
Our experiments showed that the both-goals method manages to detect incompatible goals but still performs similarly to independent (we discuss this result in more detail in Appendix \ref{app:bothgoals}).
 This suggests that the reason why independent fails is the large number of infeasible episodes and agrees with our observation that, when we decrease the number of goals in the environment, which dramatically reduces the probability of infeasible episodes, independent reaches the performance of centralized (see analysis in the environment with 3 landmarks in Appendix \ref{app:3_landmarks}).
{\color{revise} A similar behavior to both-goals is observed for the CTDE baseline, which achieves a slightly better performance.
As we explain in our analysis of this method in Appendix \ref{app:CTDE}, 
including both goals in the value function enabled both CTDE and both-goals to detect infeasible episodes.
Under CTDE, agents also exhibited more intra-episode adaptation, which may explain their superior performance.
The baseline with the recurrent policy, analyzed in Appendix \ref{app:recurrent}, faces the same limitation and is more sensitive to the noise introduced by infeasible episodes compared to feedforward policies.
}

We should note that alignment is not sufficient for acting optimally in our environment as, even if both agents choose the same cooperative goal they still need to coordinate on who goes where.
How can they do so with perfect success rate?
We hypothesize that the agents will find it challenging to adapt to the other's behavior due to the high level of partial observability in the environment:
without a recurrent policy and without observing the direction an agent is moving to, inferring the sub-goal pursued by the other is difficult.
Instead, a specialization strategy where the two agents reach an agreement during training on who goes where
(e.g. one agent always goes to the left-most landmark and the other to the rightmost) requires less effort.
To detect this behavior, we search for specialization, i.e., policies that, when assigned with a cooperative goal during evaluation, are biased to one of its landmarks.
We quantitavely measure specialization as the ratio of the episodes in which the agent
went to its preferred landmark when following a cooperative goal.
For example, if for goal $[101000]$ an agent went 7 times to $[100000]$ and 3 times to $[00100]$ this score would be 0.7.
We observed that specialization correlates with alignment: independent specializes by $\textstyle 0.72 \pm 0.0452$, $50\%$-align by $\textstyle 0.8066 \pm 0.0537$ and centralized by $ \textstyle 0.92 \pm 0.083$ (see Figure \ref{fig:special} in Appendix \ref{app:special} for an illustration of these results)
.


\subsection{Learning to align goals}\label{sec:game_results}
We have established that alignment is an efficient strategy for solving the \probname.
To investigate whether it can be achieved without introducing centralization, we now turn to the evaluation of our proposed method for coordinating goals through communication, the \algname, that we described in Section \ref{sec:game}.
We observe that,in Figure~\ref{fig:performance}, the evaluation reward for the \algname at the end of training is $\textstyle 0.9144 \pm 0.0044$.
We also observe that early in training (time step $\textstyle 66\cdot10^5$) the \algname collects less rewards than centralized. 
Similarly for the episode length, the \algname is initially slower than centralized but at the end of training reaches its speed and surpasses independent and $50\%$-align. Next, we take a deeper look at its dynamics to understand these behaviors. In particular, we study the update matrices at early and later stages of training to understand why performance starts off bad but then reaches the optimal value.
\begin{figure}
    \centering
    \includegraphics[width=0.85\textwidth]{figures/results/matrices}
    \caption{Evolution of the matrices of the \algname early (left), in the middle (middle) and at the end of training (right): \textmd{Rows correspond to goals, with individual goals assigned to the first 6 rows, columns correspond to messages and the intensity of a cell indicates the confidence in a goal-message association.} The green arrow highlights communication that leads to goal alignment (both agents pursue goal $[001010]$) and the red communication that leads to the "risky follower" behavior (agent 0 pursues goal $[000001]$ while agent 1 pursues goal $[010001]$).}
\label{fig:matrices}
\end{figure}

In Figure~\ref{fig:matrices}, we visualize the matrices for a simulation that differs from the one in Figure~\ref{fig:performance} only in that $\beta$ is increased from 2 to 4.
As we discuss in Appendix \ref{app:multiplier}, increasing $\beta$ does not affect performance but leads to interesting emerging behaviors that our study of the matrices can reveal. 
As we described in Section \ref{sec:game}, each agent is equipped with a matrix mapping messages to goals and updates its cells to maximize its individual rewards during training episodes. The rows of the matrices correspond to goals and the columns to messages.
We make the convention here of plotting the individual goals first, so the first 6 rows correspond to individual and the following 15 to cooperative goals.
We have set the message size to a slightly higher value than the number of goals, i.e., $M=30$.
As we show in Appendix \ref{app:messages} having more messages than goals facilitates training by decreasing the probability that the matrix updating will get stuck.
Rows and columns where we can find a single cell with higher intensity than others indicate a converged goal-message association.
We can detect alignment by tracing if the goal-message associations of the two agents agree.
We observe that, early in training, the agents have low confidence for most associations and alignment has not been achieved. By the middle of training, however, the two matrices are almost identical (follow the green arrow for an example of communication leading to aligned goals). 
Looking at the matrices in Figure \ref{fig:matrices} we can see that not all goals are aligned.
This is because in some cases (see the red line for an example) the leader samples an individual goal that the follower interprets as a cooperative one.
This "risky" behavior is useful for the follower, as cooperative goals are more rewarding than individual ones.
If the message received by the follower convinces it that the leader is pursuing an individual goal, the follower might have interest to pursue a cooperative goal compatible with it, which will lead to maximum reward. In a daily life analogy, if your housemate tells you they will buy pasta for tonight (their individual goal), you may buy pasta sauce (a cooperative goal based on your expectation that the other will fulfill its individual goal) instead of rice (a different individual goal).

\begin{figure}
\begin{minipage}[t]{0.9\textwidth} 
\vspace{0cm}
    \centering
    \includegraphics[width=0.72\textwidth]{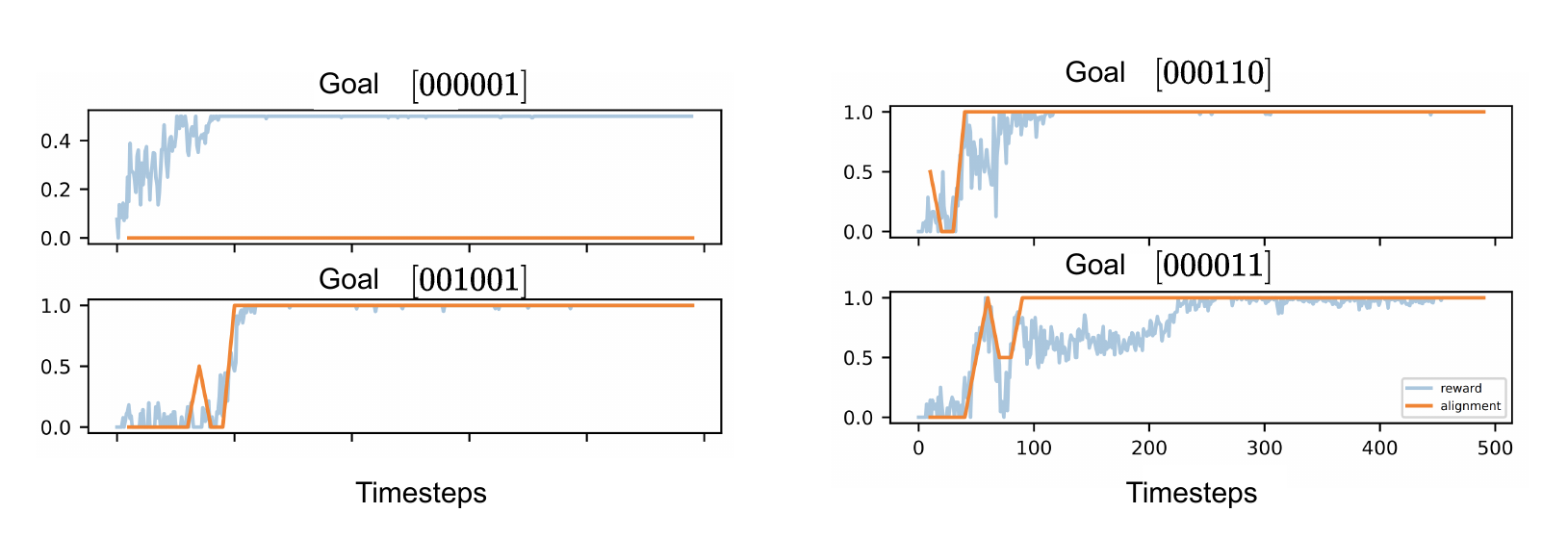}
    \caption{Co-evolution of alignment and rewards for different goals during training using the \algname. The top-left plot corresponds to an individual goal and the rest to cooperative ones.}
    \label{fig:coev} 
\end{minipage}
\end{figure}

A challenging feature of the \algname is that the matrices and polices are updated simultaneously.
This can lead to a chicken-and-egg problem: the matrix updates may fail even if the goal-message association is correct because the policy has not managed to solve a goal. Or the policy may struggle to solve goals because of bad goal-message associations that lead to episodes infeasible to solve.
In Figure \ref{fig:coev} we monitor the co-evolution of alignment and rewards during training for a random subset of the goals. We observe that, for cooperative goals, rewards and alignment are highly correlated with improvements in one driving improvements in the other, while, for individual goals, rewards are maximized without requiring alignment.

\section{Discussion}\label{sec:discuss}
We present a new problem for formalizing intrinsically-motivated multi-agent goal exploration in a decentralized
training paradigm, \probname, and propose an algorithm for solving it, the \algname.
We empirically observe that shared intentionality, which we measure as alignment of cooperative goals during training,
plays an important role in a group's ability to solve a wide diversity of tasks.
Aligned agents do not only get the highest rewards but also do so quickly.
We also show that, under the \algname, alignment emerges without being explicitly rewarded and groups reach equal performance to a centralized setting that guarantees alignment. We observed that groups with higher alignment solve the tasks by specializing
instead of monitoring and adapting to others, which, as has been observed in previous MARL studies~\citep{Ndousse2021EmergentSL}, is a behavior challenging to emerge unless explicitly rewarded.
{\color{revise}
We have adopted a descriptive rather than normative approach, common in the study of open-ended learning~\citep{Baker2020Emergent,chan2020LeniaExpandedUniverse}. Our aim was to get groups of agents that learn a maximally diverse behavioral repertoire, but observed the emergence of behaviors such as the risky follower.
Whether such behaviors are desirable or not, depends on the application at hand. 
}

Our study of the \algname is limited to populations of two agents and discrete message and goal spaces.
Extending it to larger groups is important for scaling up its applicability.
We hypothesize that in such settings specialization will no longer lead to optimal performance and that the goal-conditioned policy
will need to be extended by conditioning it on messages and introducing recurrency to equip agents with memory \citep{duan2016rl}.
Also, having shown that increasing environmental complexity increases the importance of alignment (by comparing environments with different numbers of landmakrs), we believe that an interesting extension of this work would be to test our approach in a more complex, multi-agent environment like
Grafter \citep{grafter}.
To extend the \algname to continuous message and goal spaces, we can adopt approaches based on energy-based models employed
in previous works \citep{https://doi.org/10.48550/arxiv.2210.06468}.
Finally, while our empirical study considers a pre-defined goal space, we should note that this is not necessary for autotelic agents who can in general learn their own goal representation
\citep{colas2022autotelic}.
We envision  studies of the Goal-coordination game where both goals and messages emerge (to study for example language evolution \citep{daganCoevolutionLanguageAgents2021,moulin-frier2021MultiAgentReinforcementLearning}).

We believe that the \probname, can be of interest in real-world scenarios such as robotics for disaster rescue or extraterrestrial exploration.
It allows to consider a population of goal-conditioned RL agents that learn how to achieve a wide diversity of
cooperative tasks in a fully-autonomous manner.
In this way, a user could place agents (simulated or robotics) in some
environment and let them interact without any supervision for a period of time.
At the end of this training phase, the agent population will have autonomously learned how to achieve diverse individual
and collaborative goals without any supervision and a human user will be able to benefit from these acquired skills.

\paragraph{Acknowledgements}
This research was partially funded by the French National Research Agency (\url{https://anr.fr/}, project ECOCURL, Grant ANR-20-CE23-0006). This work also benefited from access to the HPC resources of IDRIS under the allocation 2020-[A0091011996] made by GENCI. 

\bibliography{collas2023_conference}

\begin{thebibliography}{58}
\providecommand{\natexlab}[1]{#1}
\providecommand{\url}[1]{\texttt{#1}}
\expandafter\ifx\csname urlstyle\endcsname\relax
  \providecommand{\doi}[1]{doi: #1}\else
  \providecommand{\doi}{doi: \begingroup \urlstyle{rm}\Url}\fi

\bibitem[Achiam \& Sastry(2017)Achiam and
  Sastry]{achiam2017SurpriseBasedIntrinsicMotivation}
Joshua Achiam and Shankar Sastry.
\newblock Surprise-{{Based Intrinsic Motivation}} for {{Deep Reinforcement
  Learning}}, March 2017.

\bibitem[Andrychowicz et~al.(2017)Andrychowicz, Wolski, Ray, Schneider, Fong,
  Welinder, McGrew, Tobin, Pieter~Abbeel, and
  Zaremba]{andrychowicz2017hindsight}
Marcin Andrychowicz, Filip Wolski, Alex Ray, Jonas Schneider, Rachel Fong,
  Peter Welinder, Bob McGrew, Josh Tobin, OpenAI Pieter~Abbeel, and Wojciech
  Zaremba.
\newblock Hindsight experience replay.
\newblock \emph{Advances in neural information processing systems}, 30, 2017.

\bibitem[Andrychowicz et~al.(2020)Andrychowicz, Raichuk, Sta{\'n}czyk, Orsini,
  Girgin, Marinier, Hussenot, Geist, Pietquin, Michalski,
  et~al.]{andrychowicz2020matters}
Marcin Andrychowicz, Anton Raichuk, Piotr Sta{\'n}czyk, Manu Orsini, Sertan
  Girgin, Rapha{\"e}l Marinier, Leonard Hussenot, Matthieu Geist, Olivier
  Pietquin, Marcin Michalski, et~al.
\newblock What matters for on-policy deep actor-critic methods? a large-scale
  study.
\newblock In \emph{International conference on learning representations}, 2020.

\bibitem[Baker et~al.(2020)Baker, Kanitscheider, Markov, Wu, Powell, McGrew,
  and Mordatch]{Baker2020Emergent}
Bowen Baker, Ingmar Kanitscheider, Todor Markov, Yi~Wu, Glenn Powell, Bob
  McGrew, and Igor Mordatch.
\newblock Emergent tool use from multi-agent autocurricula.
\newblock In \emph{International Conference on Learning Representations}, 2020.
\newblock URL \url{https://openreview.net/forum?id=SkxpxJBKwS}.

\bibitem[Bamford(2022)]{grafter}
Chris Bamford.
\newblock Grafter.
\newblock \url{https://github.com/GriddlyAI/grafter}, 2022.

\bibitem[Berlyne(1966)]{berlyne1966curiosity}
Daniel~E Berlyne.
\newblock Curiosity and exploration: Animals spend much of their time seeking
  stimuli whose significance raises problems for psychology.
\newblock \emph{Science}, 153\penalty0 (3731):\penalty0 25--33, 1966.

\bibitem[Bernstein et~al.(2013)Bernstein, Zilberstein, and
  Immerman]{bernstein2013ComplexityDecentralizedControl}
Daniel~S. Bernstein, Shlomo Zilberstein, and Neil Immerman.
\newblock The {{Complexity}} of {{Decentralized Control}} of {{Markov Decision
  Processes}}, January 2013.

\bibitem[Blaes et~al.(2020)Blaes, Pogan{\v c}i{\'c}, Zhu, and
  Martius]{blaes2020ControlWhatYou}
Sebastian Blaes, Marin~Vlastelica Pogan{\v c}i{\'c}, Jia-Jie Zhu, and Georg
  Martius.
\newblock Control {{What You Can}}: {{Intrinsically Motivated Task-Planning
  Agent}}, January 2020.

\bibitem[Chan(2020)]{chan2020LeniaExpandedUniverse}
Bert Wang-Chak Chan.
\newblock Lenia and {{Expanded Universe}}.
\newblock In \emph{The 2020 {{Conference}} on {{Artificial Life}}}, pp.\
  221--229, 2020.
\newblock \doi{10.1162/isal_a_00297}.

\bibitem[Chentanez et~al.(2004)Chentanez, Barto, and Singh]{NIPS2004_4be5a36c}
Nuttapong Chentanez, Andrew Barto, and Satinder Singh.
\newblock Intrinsically motivated reinforcement learning.
\newblock In L.~Saul, Y.~Weiss, and L.~Bottou (eds.), \emph{Advances in Neural
  Information Processing Systems}, volume~17. MIT Press, 2004.
\newblock URL
  \url{https://proceedings.neurips.cc/paper/2004/file/4be5a36cbaca8ab9d2066debfe4e65c1-Paper.pdf}.

\bibitem[Claus \& Boutilier(1998)Claus and Boutilier]{claus1998dynamics}
Caroline Claus and Craig Boutilier.
\newblock The dynamics of reinforcement learning in cooperative multiagent
  systems.
\newblock \emph{AAAI/IAAI}, 1998\penalty0 (746-752):\penalty0 2, 1998.

\bibitem[Colas et~al.(2019)Colas, Fournier, Sigaud, Chetouani, and
  Oudeyer]{https://doi.org/10.48550/arxiv.1810.06284}
C{\'e}dric Colas, Pierre Fournier, Olivier Sigaud, Mohamed Chetouani, and
  Pierre-Yves Oudeyer.
\newblock {CURIOUS: Intrinsically Motivated Modular Multi-Goal Reinforcement
  Learning}.
\newblock In \emph{{ICML 2019 - Thirty-sixth International Conference on
  Machine Learning}}, Long Beach, United States, June 2019.
\newblock URL \url{https://hal.science/hal-01934921}.

\bibitem[Colas et~al.(2020)Colas, Karch, Lair, Dussoux, Moulin-Frier, Dominey,
  and Oudeyer]{colas2020language}
C{\'e}dric Colas, Tristan Karch, Nicolas Lair, Jean-Michel Dussoux, Cl{\'e}ment
  Moulin-Frier, Peter Dominey, and Pierre-Yves Oudeyer.
\newblock Language as a cognitive tool to imagine goals in curiosity driven
  exploration.
\newblock \emph{Advances in Neural Information Processing Systems},
  33:\penalty0 3761--3774, 2020.

\bibitem[Colas et~al.(2022)Colas, Karch, Sigaud, and
  Oudeyer]{colas2022autotelic}
C{\'e}dric Colas, Tristan Karch, Olivier Sigaud, and Pierre-Yves Oudeyer.
\newblock Autotelic agents with intrinsically motivated goal-conditioned
  reinforcement learning: a short survey.
\newblock \emph{Journal of Artificial Intelligence Research}, 74:\penalty0
  1159--1199, 2022.

\bibitem[Dagan et~al.(2021)Dagan, Hupkes, and
  Bruni]{daganCoevolutionLanguageAgents2021}
Gautier Dagan, Dieuwke Hupkes, and Elia Bruni.
\newblock Co-evolution of language and agents in referential games.
\newblock \emph{arXiv:2001.03361 [cs]}, January 2021.

\bibitem[Duan et~al.(2016)Duan, Schulman, Chen, Bartlett, Sutskever, and
  Abbeel]{duan2016rl}
Yan Duan, John Schulman, Xi~Chen, Peter~L Bartlett, Ilya Sutskever, and Pieter
  Abbeel.
\newblock Fast reinforcement learning via slow reinforcement learning.
\newblock \emph{arXiv preprint arXiv:1611.02779}, 2016.

\bibitem[Foerster et~al.(2017)Foerster, Farquhar, Afouras, Nardelli, and
  Whiteson]{foerster2017CounterfactualMultiAgentPolicy}
Jakob Foerster, Gregory Farquhar, Triantafyllos Afouras, Nantas Nardelli, and
  Shimon Whiteson.
\newblock Counterfactual {{Multi-Agent Policy Gradients}}, December 2017.

\bibitem[Foerster et~al.(2016)Foerster, Assael, {de Freitas}, and
  Whiteson]{foersterLearningCommunicateDeep2016}
Jakob~N. Foerster, Yannis~M. Assael, Nando {de Freitas}, and Shimon Whiteson.
\newblock Learning to {{Communicate}} with {{Deep Multi-Agent Reinforcement
  Learning}}.
\newblock \emph{arXiv:1605.06676 [cs]}, May 2016.

\bibitem[Garcia~Ortiz et~al.(2021)Garcia~Ortiz, Jankovics, Caselles-Dupre, and
  Annabi]{Simple-Playgrounds}
Michael Garcia~Ortiz, Vince Jankovics, Hugo Caselles-Dupre, and Louis Annabi.
\newblock Simple-playgrounds.
\newblock \url{https://github.com/mgarciaortiz/simple-playgrounds}, 2021.

\bibitem[Gopnik et~al.(1999)Gopnik, Meltzoff, and Kuhl]{gopnik1999scientist}
Alison Gopnik, Andrew~N Meltzoff, and Patricia~K Kuhl.
\newblock \emph{The scientist in the crib: Minds, brains, and how children
  learn.}
\newblock William Morrow \& Co, 1999.

\bibitem[Hamann et~al.(2012)Hamann, Warneken, and
  Tomasello]{hamann2012children}
Katharina Hamann, Felix Warneken, and Michael Tomasello.
\newblock Children’s developing commitments to joint goals.
\newblock \emph{Child development}, 83\penalty0 (1):\penalty0 137--145, 2012.

\bibitem[Jaderberg et~al.(2019)Jaderberg, Czarnecki, Dunning, Marris, Lever,
  Castañeda, Beattie, Rabinowitz, Morcos, Ruderman, Sonnerat, Green, Deason,
  Leibo, Silver, Hassabis, Kavukcuoglu, and
  Graepel]{doi:10.1126/science.aau6249}
Max Jaderberg, Wojciech~M. Czarnecki, Iain Dunning, Luke Marris, Guy Lever,
  Antonio~Garcia Castañeda, Charles Beattie, Neil~C. Rabinowitz, Ari~S.
  Morcos, Avraham Ruderman, Nicolas Sonnerat, Tim Green, Louise Deason, Joel~Z.
  Leibo, David Silver, Demis Hassabis, Koray Kavukcuoglu, and Thore Graepel.
\newblock Human-level performance in 3d multiplayer games with population-based
  reinforcement learning.
\newblock \emph{Science}, 364\penalty0 (6443):\penalty0 859--865, 2019.
\newblock \doi{10.1126/science.aau6249}.
\newblock URL \url{https://www.science.org/doi/abs/10.1126/science.aau6249}.

\bibitem[Jaques et~al.(2019)Jaques, Lazaridou, Hughes, Gulcehre, Ortega,
  Strouse, Leibo, and De~Freitas]{jaques2019social}
Natasha Jaques, Angeliki Lazaridou, Edward Hughes, Caglar Gulcehre, Pedro
  Ortega, DJ~Strouse, Joel~Z Leibo, and Nando De~Freitas.
\newblock Social influence as intrinsic motivation for multi-agent deep
  reinforcement learning.
\newblock In \emph{International conference on machine learning}, pp.\
  3040--3049. PMLR, 2019.

\bibitem[Kaelbling(1993)]{kaelbling1993LearningAchieveGoalsa}
L.~Kaelbling.
\newblock Learning to {{Achieve Goals}}.
\newblock In \emph{International {{Joint Conference}} on {{Artificial
  Intelligence}}}, 1993.

\bibitem[Kingma \& Ba(2014)Kingma and Ba]{kingma2014adam}
Diederik~P Kingma and Jimmy Ba.
\newblock Adam: A method for stochastic optimization.
\newblock \emph{arXiv preprint arXiv:1412.6980}, 2014.

\bibitem[Konidaris \& Barto(2009)Konidaris and
  Barto]{konidaris2009SkillDiscoveryContinuous}
George Konidaris and Andrew Barto.
\newblock Skill {{Discovery}} in {{Continuous Reinforcement Learning Domains}}
  using {{Skill Chaining}}.
\newblock In \emph{Advances in {{Neural Information Processing Systems}}},
  volume~22. {Curran Associates, Inc.}, 2009.

\bibitem[Kova{\v c} et~al.(2022)Kova{\v c}, {Laversanne-Finot}, and
  Oudeyer]{kovac2022GRIMGEPLearningProgress}
Grgur Kova{\v c}, Adrien {Laversanne-Finot}, and Pierre-Yves Oudeyer.
\newblock {{GRIMGEP}}: {{Learning Progress}} for {{Robust Goal Sampling}} in
  {{Visual Deep Reinforcement Learning}}.
\newblock \emph{IEEE Transactions on Cognitive and Developmental Systems}, pp.\
   1--1, 2022.
\newblock ISSN 2379-8939.
\newblock \doi{10.1109/TCDS.2022.3216911}.

\bibitem[Lazaridou \& Baroni(2020)Lazaridou and
  Baroni]{lazaridouEmergentMultiAgentCommunication2020}
Angeliki Lazaridou and Marco Baroni.
\newblock Emergent {{Multi-Agent Communication}} in the {{Deep Learning Era}}.
\newblock \emph{arXiv:2006.02419 [cs]}, July 2020.

\bibitem[Lemesle et~al.(2022)Lemesle, Karch, Laroche, Moulin-Frier, and
  Oudeyer]{https://doi.org/10.48550/arxiv.2210.06468}
Yoann Lemesle, Tristan Karch, Romain Laroche, Cl{\'e}ment Moulin-Frier, and
  Pierre-Yves Oudeyer.
\newblock Emergence of shared sensory-motor graphical language from visual
  input, 2022.
\newblock URL \url{https://arxiv.org/abs/2210.06468}.

\bibitem[Liang et~al.(2018)Liang, Liaw, Nishihara, Moritz, Fox, Goldberg,
  Gonzalez, Jordan, and Stoica]{pmlr-v80-liang18b}
Eric Liang, Richard Liaw, Robert Nishihara, Philipp Moritz, Roy Fox, Ken
  Goldberg, Joseph Gonzalez, Michael Jordan, and Ion Stoica.
\newblock {RL}lib: Abstractions for distributed reinforcement learning.
\newblock In Jennifer Dy and Andreas Krause (eds.), \emph{Proceedings of the
  35th International Conference on Machine Learning}, volume~80 of
  \emph{Proceedings of Machine Learning Research}, pp.\  3053--3062. PMLR,
  10--15 Jul 2018.
\newblock URL \url{https://proceedings.mlr.press/v80/liang18b.html}.

\bibitem[Littman(1994)]{littman1994MarkovGamesFramework}
Michael~L. Littman.
\newblock Markov games as a framework for multi-agent reinforcement learning.
\newblock In \emph{Proceedings of the {{Eleventh International Conference}} on
  {{International Conference}} on {{Machine Learning}}}, {{ICML}}'94, pp.\
  157--163, {San Francisco, CA, USA}, July 1994. {Morgan Kaufmann Publishers
  Inc.}
\newblock ISBN 978-1-55860-335-6.

\bibitem[Liu et~al.(2022)Liu, Zhu, and
  Zhang]{liu2022GoalConditionedReinforcementLearning}
Minghuan Liu, Menghui Zhu, and Weinan Zhang.
\newblock Goal-{{Conditioned Reinforcement Learning}}: {{Problems}} and
  {{Solutions}}, September 2022.

\bibitem[Lowe et~al.(2020)Lowe, Wu, Tamar, Harb, Abbeel, and
  Mordatch]{loweMultiAgentActorCriticMixed2020}
Ryan Lowe, Yi~Wu, Aviv Tamar, Jean Harb, Pieter Abbeel, and Igor Mordatch.
\newblock Multi-{{Agent Actor-Critic}} for {{Mixed Cooperative-Competitive
  Environments}}.
\newblock \emph{arXiv:1706.02275 [cs]}, March 2020.

\bibitem[Mordatch \& Abbeel(2018)Mordatch and Abbeel]{mordatch2018emergence}
Igor Mordatch and Pieter Abbeel.
\newblock Emergence of grounded compositional language in multi-agent
  populations.
\newblock In \emph{Proceedings of the AAAI Conference on Artificial
  Intelligence}, volume~32, 2018.

\bibitem[Moulin-Frier \& Oudeyer(2021)Moulin-Frier and
  Oudeyer]{moulin-frier2021MultiAgentReinforcementLearning}
Clément Moulin-Frier and Pierre-Yves Oudeyer.
\newblock Multi-{Agent} {Reinforcement} {Learning} as a {Computational} {Tool}
  for {Language} {Evolution} {Research}: {Historical} {Context} and {Future}
  {Challenges}.
\newblock In \emph{Challenges and {Opportunities} for {Multi}-{Agent}
  {Reinforcement} {Learning} ({COMARL}), {AAAI} {Spring} {Symposium} {Series},
  {Stanford} {University}, {Palo} {Alto}, {California}, {USA}}, 2021.
\newblock URL \url{http://arxiv.org/abs/2002.08878}.
\newblock tex.ids= moulin-frier2020MultiAgentReinforcementLearning arXiv:
  2002.08878.

\bibitem[Nair et~al.(2018)Nair, Pong, Dalal, Bahl, Lin, and
  Levine]{nair2018VisualReinforcementLearning}
Ashvin~V Nair, Vitchyr Pong, Murtaza Dalal, Shikhar Bahl, Steven Lin, and
  Sergey Levine.
\newblock Visual {{Reinforcement Learning}} with {{Imagined Goals}}.
\newblock In \emph{Advances in {{Neural Information Processing Systems}}},
  volume~31. {Curran Associates, Inc.}, 2018.

\bibitem[Ndousse et~al.(2021)Ndousse, Eck, Levine, and
  Jaques]{Ndousse2021EmergentSL}
Kamal Ndousse, Douglas Eck, Sergey Levine, and Natasha Jaques.
\newblock Emergent social learning via multi-agent reinforcement learning.
\newblock In \emph{ICML}, 2021.

\bibitem[Nguyen \& Oudeyer(2014{\natexlab{a}})Nguyen and
  Oudeyer]{nguyen2014SociallyGuidedIntrinsic}
Sao~Mai Nguyen and Pierre-Yves Oudeyer.
\newblock Socially guided intrinsic motivation for robot learning of motor
  skills.
\newblock \emph{Autonomous Robots}, 36\penalty0 (3):\penalty0 273--294, March
  2014{\natexlab{a}}.
\newblock ISSN 0929-5593, 1573-7527.
\newblock \doi{10.1007/s10514-013-9339-y}.

\bibitem[Nguyen \& Oudeyer(2014{\natexlab{b}})Nguyen and
  Oudeyer]{nguyen2014socially}
Sao~Mai Nguyen and Pierre-Yves Oudeyer.
\newblock Socially guided intrinsic motivation for robot learning of motor
  skills.
\newblock \emph{Autonomous Robots}, 36\penalty0 (3):\penalty0 273--294,
  2014{\natexlab{b}}.

\bibitem[Oudeyer \& Kaplan(2009)Oudeyer and Kaplan]{oudeyer2009intrinsic}
Pierre-Yves Oudeyer and Frederic Kaplan.
\newblock What is intrinsic motivation? a typology of computational approaches.
\newblock \emph{Frontiers in neurorobotics}, pp.\ ~6, 2009.

\bibitem[Parisi et~al.(2019)Parisi, Kemker, Part, Kanan, and
  Wermter]{parisi2019ContinualLifelongLearning}
German~I. Parisi, Ronald Kemker, Jose~L. Part, Christopher Kanan, and Stefan
  Wermter.
\newblock Continual lifelong learning with neural networks: {{A}} review.
\newblock \emph{Neural Networks}, 113:\penalty0 54--71, May 2019.
\newblock ISSN 0893-6080.
\newblock \doi{10.1016/j.neunet.2019.01.012}.

\bibitem[Paszke et~al.(2019)Paszke, Gross, Massa, Lerer, Bradbury, Chanan,
  Killeen, Lin, Gimelshein, Antiga, Desmaison, Kopf, Yang, DeVito, Raison,
  Tejani, Chilamkurthy, Steiner, Fang, Bai, and Chintala]{NEURIPS2019_9015}
Adam Paszke, Sam Gross, Francisco Massa, Adam Lerer, James Bradbury, Gregory
  Chanan, Trevor Killeen, Zeming Lin, Natalia Gimelshein, Luca Antiga, Alban
  Desmaison, Andreas Kopf, Edward Yang, Zachary DeVito, Martin Raison, Alykhan
  Tejani, Sasank Chilamkurthy, Benoit Steiner, Lu~Fang, Junjie Bai, and Soumith
  Chintala.
\newblock Pytorch: An imperative style, high-performance deep learning library.
\newblock In H.~Wallach, H.~Larochelle, A.~Beygelzimer, F.~d\textquotesingle
  Alch\'{e}-Buc, E.~Fox, and R.~Garnett (eds.), \emph{Advances in Neural
  Information Processing Systems 32}, pp.\  8024--8035. Curran Associates,
  Inc., 2019.
\newblock URL
  \url{http://papers.neurips.cc/paper/9015-pytorch-an-imperative-style-high-performance-deep-learning-library.pdf}.

\bibitem[Pathak et~al.(2017)Pathak, Agrawal, Efros, and
  Darrell]{pathak2017CuriositydrivenExplorationSelfsupervised}
Deepak Pathak, Pulkit Agrawal, Alexei~A. Efros, and Trevor Darrell.
\newblock Curiosity-driven {{Exploration}} by {{Self-supervised Prediction}},
  May 2017.

\bibitem[Pong et~al.(2020)Pong, Dalal, Lin, Nair, Bahl, and
  Levine]{pong2020SkewFitStateCoveringSelfSupervised}
Vitchyr~H. Pong, Murtaza Dalal, Steven Lin, Ashvin Nair, Shikhar Bahl, and
  Sergey Levine.
\newblock Skew-{{Fit}}: {{State-Covering Self-Supervised Reinforcement
  Learning}}, August 2020.

\bibitem[Rubin et~al.(1978)Rubin, Watson, and Jambor]{rubin1978free}
Kenneth~H Rubin, Kathryn~S Watson, and Thomas~W Jambor.
\newblock Free-play behaviors in preschool and kindergarten children.
\newblock \emph{Child development}, pp.\  534--536, 1978.

\bibitem[Schaul et~al.(2015)Schaul, Horgan, Gregor, and
  Silver]{schaul2015UniversalValueFunction}
Tom Schaul, Daniel Horgan, Karol Gregor, and David Silver.
\newblock Universal {{Value Function Approximators}}.
\newblock In \emph{Proceedings of the 32nd {{International Conference}} on
  {{Machine Learning}}}, pp.\  1312--1320. {PMLR}, June 2015.

\bibitem[Schmidhuber(2009)]{schmidhuber2009DrivenCompressionProgress}
Juergen Schmidhuber.
\newblock Driven by {{Compression Progress}}: {{A Simple Principle Explains
  Essential Aspects}} of {{Subjective Beauty}}, {{Novelty}}, {{Surprise}},
  {{Interestingness}}, {{Attention}}, {{Curiosity}}, {{Creativity}}, {{Art}},
  {{Science}}, {{Music}}, {{Jokes}}, April 2009.

\bibitem[Schulman et~al.(2015)Schulman, Moritz, Levine, Jordan, and
  Abbeel]{schulman2015high}
John Schulman, Philipp Moritz, Sergey Levine, Michael Jordan, and Pieter
  Abbeel.
\newblock High-dimensional continuous control using generalized advantage
  estimation.
\newblock \emph{arXiv preprint arXiv:1506.02438}, 2015.

\bibitem[Skyrms(2001)]{skyrms2001StagHunt}
Brian Skyrms.
\newblock The {{Stag Hunt}}.
\newblock \emph{Proceedings and Addresses of the American Philosophical
  Association}, 75\penalty0 (2):\penalty0 31--41, 2001.
\newblock ISSN 0065-972X.
\newblock \doi{10.2307/3218711}.

\bibitem[Steels(2015)]{Steels2015}
Luc~L. Steels.
\newblock \emph{The {Talking} {Heads} experiment}.
\newblock Number~1 in Computational Models of Language Evolution. Language
  Science Press, Berlin, 2015.
\newblock \doi{10.17169/FUDOCS_document_000000022455}.

\bibitem[Sutton et~al.(2011)Sutton, Modayil, Delp, Degris, Pilarski, White, and
  Precup]{sutton2011HordeScalableRealtime}
R.~Sutton, Joseph Modayil, M.~Delp, T.~Degris, P.~Pilarski, Adam White, and
  Doina Precup.
\newblock Horde: A scalable real-time architecture for learning knowledge from
  unsupervised sensorimotor interaction.
\newblock In \emph{Adaptive {{Agents}} and {{Multi-Agent Systems}}}, May 2011.

\bibitem[Sutton \& Barto(2018)Sutton and Barto]{sutton2018reinforcement}
Richard~S Sutton and Andrew~G Barto.
\newblock \emph{Reinforcement learning: An introduction}.
\newblock MIT press, 2018.

\bibitem[Sutton et~al.(1999)Sutton, Precup, and
  Singh]{sutton1999MDPsSemiMDPsFramework}
Richard~S. Sutton, Doina Precup, and Satinder Singh.
\newblock Between {{MDPs}} and semi-{{MDPs}}: {{A}} framework for temporal
  abstraction in reinforcement learning.
\newblock \emph{Artificial Intelligence}, 112\penalty0 (1):\penalty0 181--211,
  August 1999.
\newblock ISSN 0004-3702.
\newblock \doi{10.1016/S0004-3702(99)00052-1}.

\bibitem[Tomasello \& Carpenter(2007)Tomasello and
  Carpenter]{tomasello2007shared}
Michael Tomasello and Malinda Carpenter.
\newblock Shared intentionality.
\newblock \emph{Developmental science}, 10\penalty0 (1):\penalty0 121--125,
  2007.

\bibitem[Vinyals et~al.(2017)Vinyals, Ewalds, Bartunov, Georgiev, Vezhnevets,
  Yeo, Makhzani, Küttler, Agapiou, Schrittwieser, Quan, Gaffney, Petersen,
  Simonyan, Schaul, van Hasselt, Silver, Lillicrap, Calderone, Keet, Brunasso,
  Lawrence, Ekermo, Repp, and Tsing]{https://doi.org/10.48550/arxiv.1708.04782}
Oriol Vinyals, Timo Ewalds, Sergey Bartunov, Petko Georgiev, Alexander~Sasha
  Vezhnevets, Michelle Yeo, Alireza Makhzani, Heinrich Küttler, John Agapiou,
  Julian Schrittwieser, John Quan, Stephen Gaffney, Stig Petersen, Karen
  Simonyan, Tom Schaul, Hado van Hasselt, David Silver, Timothy Lillicrap,
  Kevin Calderone, Paul Keet, Anthony Brunasso, David Lawrence, Anders Ekermo,
  Jacob Repp, and Rodney Tsing.
\newblock Starcraft ii: A new challenge for reinforcement learning, 2017.
\newblock URL \url{https://arxiv.org/abs/1708.04782}.

\bibitem[Wang et~al.(2020)Wang, Kew, Lee, Lee, Zhang, Ichter, Tan, and
  Faust]{wangModelbasedReinforcementLearning2020}
Rose~E. Wang, J.~Chase Kew, Dennis Lee, Tsang-Wei~Edward Lee, Tingnan Zhang,
  Brian Ichter, Jie Tan, and Aleksandra Faust.
\newblock Model-based {{Reinforcement Learning}} for {{Decentralized Multiagent
  Rendezvous}}.
\newblock \emph{arXiv:2003.06906 [cs]}, November 2020.

\bibitem[Warneken et~al.(2014)Warneken, Steinwender, Hamann, and
  Tomasello]{warneken2014young}
Felix Warneken, Jasmin Steinwender, Katharina Hamann, and Michael Tomasello.
\newblock Young children's planning in a collaborative problem-solving task.
\newblock \emph{Cognitive Development}, 31:\penalty0 48--58, 2014.

\bibitem[Yang et~al.(2020)Yang, Nakhaei, Isele, Fujimura, and Zha]{yang2018cm3}
Jiachen Yang, Alireza Nakhaei, David Isele, Kikuo Fujimura, and Hongyuan Zha.
\newblock Cm3: Cooperative multi-goal multi-stage multi-agent reinforcement
  learning.
\newblock In \emph{International Conference on Learning Representations}, 2020.
\newblock URL \url{https://openreview.net/forum?id=S1lEX04tPr}.

\end{thebibliography}
\bibliographystyle{collas2023_conference}

\cleardoublepage
\appendix

This appendix provides additional information about our set-up, implemenation details and results.
\begin{itemize}
    \item Section \ref{app:playground} describes  our navigation tasks as MDPs;
    \item Section \ref{app:hypers} provides the hyper-parameters used in the main paper;
    \item Section \ref{app:baselines} intents to clarify how the different baselines we have evaluated differ algorithmically;
    \item Section \ref{app:training} contains an empirical analysis of how the complexity of our proposed algorithms changes with task difficulty;
    \item Section \ref{app:results} contains additional results. Specifically, Section \ref{app:3_landmarks} shows the effect of environmental complexity, Section \ref{app:messages} shows the effect of message size in the \algname, Section \ref{app:recurrent} examines the usefulness of recurrent policies, Section \ref{app:LP} replaces random sampling with Learning Progress, Section \ref{app:coop_only} contains experiments where only cooperative goals are present in the environment, Section \ref{app:risky} presents additional information about the "risky follower policy", Section \ref{app:special} contains  specialization  results and Sections  \ref{app:bothgoals} and \ref{app:CTDE} further analyze the baseline with both goals observable and CTDE.
\end{itemize}

\section{Environment details}\label{app:playground}

The environment is implemented in Python using Simple Playgrounds~\citep{Simple-Playgrounds}. As a learning algorithm for the goal-conditioned policies we use RLlib’s PPO implementation~\citep{pmlr-v80-liang18b} and its multi-agent API with the PyTorch backend ~\citep{NEURIPS2019_9015}.

For the 3-landmarks environment the set of individual goals is $\{[001],[010], [100]\}$ and the set of cooperative goals is $\{[101],[011],$ $[110]\}$. For the 6-landmarks environment the set of individual goals is $\{[000001], [000010]$, $[000100],[001000],[010000], [100000]\}$ and the set of cooperative goals is $\{ [110000], [101000] $, $ [100100],$, $[100010] $, $[100001] , [0110000]$, $[010100], [010010] $, $ [010001]  $, $ [001100]  $, $ [001010]  $, $ [001001]$,  $[000110] , [000101]$, $[000011] $, $ [100001]  \}$

\paragraph{Observation space}
Agents are able to see each other and all the objects of the environment. We use object-centric representations, the observation vector contains the distance and the angle to each of the physical entities in the room (i.e walls, other agent, and landmarks). The order of the coordinates in the observation vector is preserved, e.g the first two coordinates are the distance and angle to the left wall. To make the navigation policy a goal-conditioned one, we concatenate the goal representation at the end of the observation vector to build the input to the networks. Observations are normalized between 0 and 1.

\paragraph{Action space}
We consider a discrete action space. Each agent is controlled by two actions: longitudinal force, and angular velocity. These actuators can take three different values: -1, 0, or 1.

\paragraph{Rewards and episodes}
Rewards are given independently to each agent conditioned on the agent's own goal. At each time step, if the goal is not fulfilled, the reward is 0, and 1 otherwise. All interactions with the environment are fully decentralized, each agent only has access to its own reward, and cannot see the reward of the others.

Once an agent gets a positive reward, the episode ends for them, i.e they cannot perform any other action but remain physically present in the environment. Episodes end either when both agents obtained their rewards or if a time limit is reached. At the beginning of an episode each agent is randomly placed inside the room, without touching any of the landmarks. The time limit in the environment was set to 250 and 500 time steps, for the 3 and 6 landmark instances respectively.

\section{Hyperparameters}\label{app:hypers}
Hyper-parameters do not vary across methods.

\paragraph{PPO} We base most of our design choices in the recommendations by \cite{andrychowicz2020matters}e:

\begin{itemize}
    \item PPO policy loss with 0.3 clipping threshold.
    \item tanh as activation function for the networks. We don't use shared layers for the value and policy networks.
    \item Generalized Advantage Estimation (GAE) \cite{schulman2015high} with $\lambda=0.9$
    \item A discount factor of $\gamma = 0.99$
    \item Adam optimizer \cite{kingma2014adam} with a learning rate of $0.0003$
\end{itemize}

From the many hyperparameters we can tune, we found that the batch size was the most relevant. After some test experiments, benchmarking results with the centralized uniform sampling baseline, we set this value to 16500 and 60000 time steps for the 3 and 6 landmarks experiments. We observed that usually a higher batch size was beneficial. For most of the hyperparameters we found that the defaults provided by RLlib were safe choices.

\begin{figure*}
\begin{minipage}{0.45\textwidth}
        \centering
    \includegraphics[width=0.7\columnwidth]{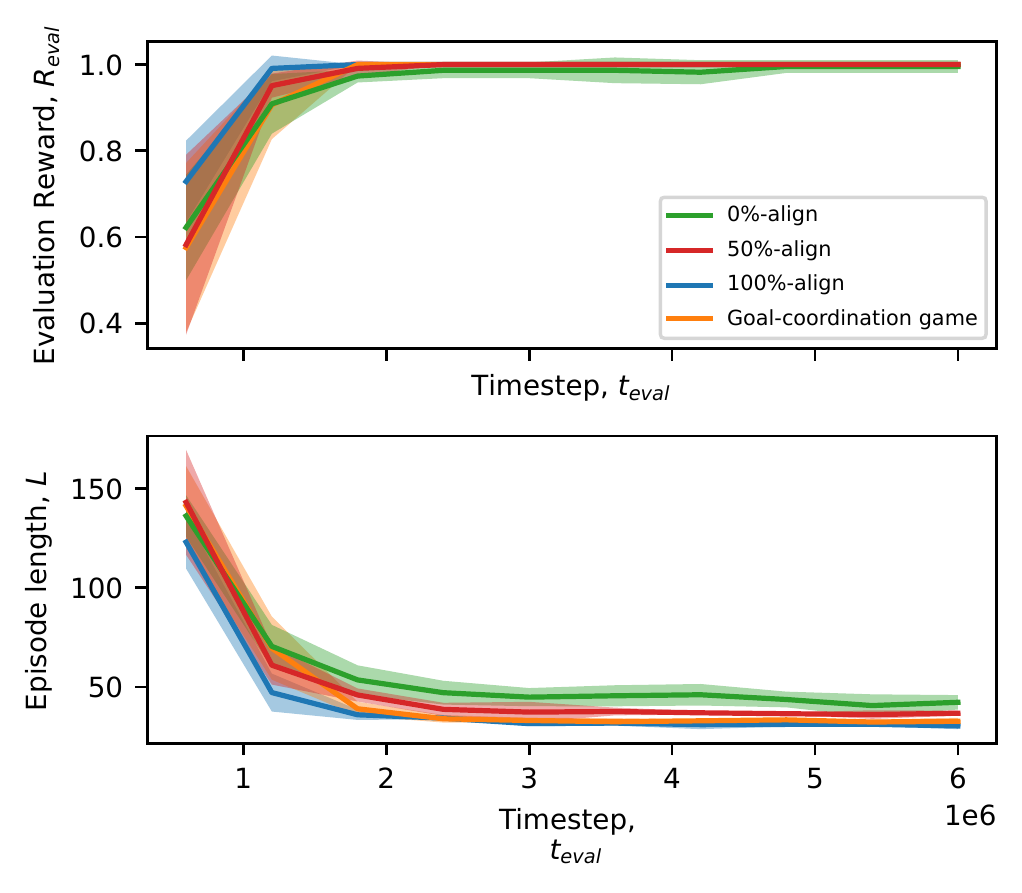}
\end{minipage}
\begin{minipage}{0.45\textwidth}
    \centering
\includegraphics[width=0.7\columnwidth]{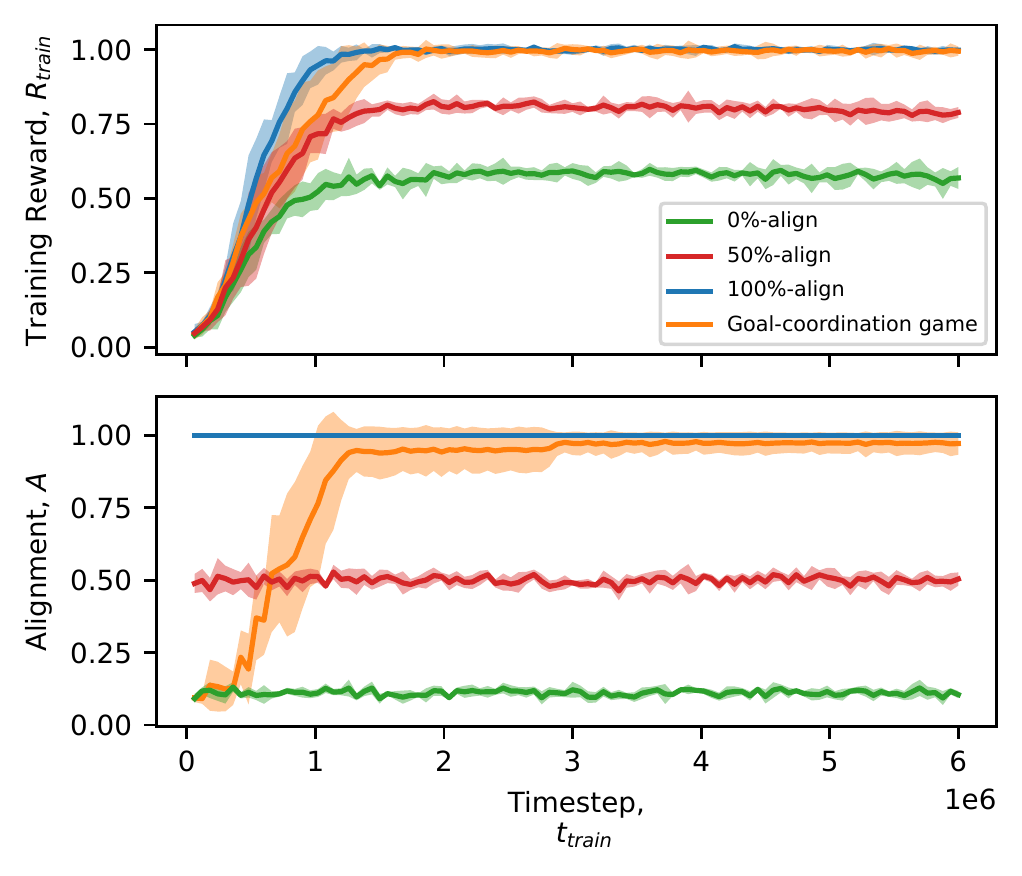}
\end{minipage}
\caption{Performance for the 3-landmarks environment during evaluation (left) and training (right) episodes for baselines exhibiting different levels of alignment and the \algname}
\label{fig:3_landmarks}
\end{figure*}
\paragraph{Goal-coordination game}
 We use a softmax of temperature $T=\frac{1}{30}$ to sample messages $m_l$ and goal $g_f$ from the matrix. The update of the matrix is made with $\alpha=0.1$ to dampen the changes of estimates of expected reward for each goal/message couple. 

\section{Illustration of baselines}\label{app:baselines}

\begin{figure*}
\centering
\includegraphics[width=0.6\columnwidth]{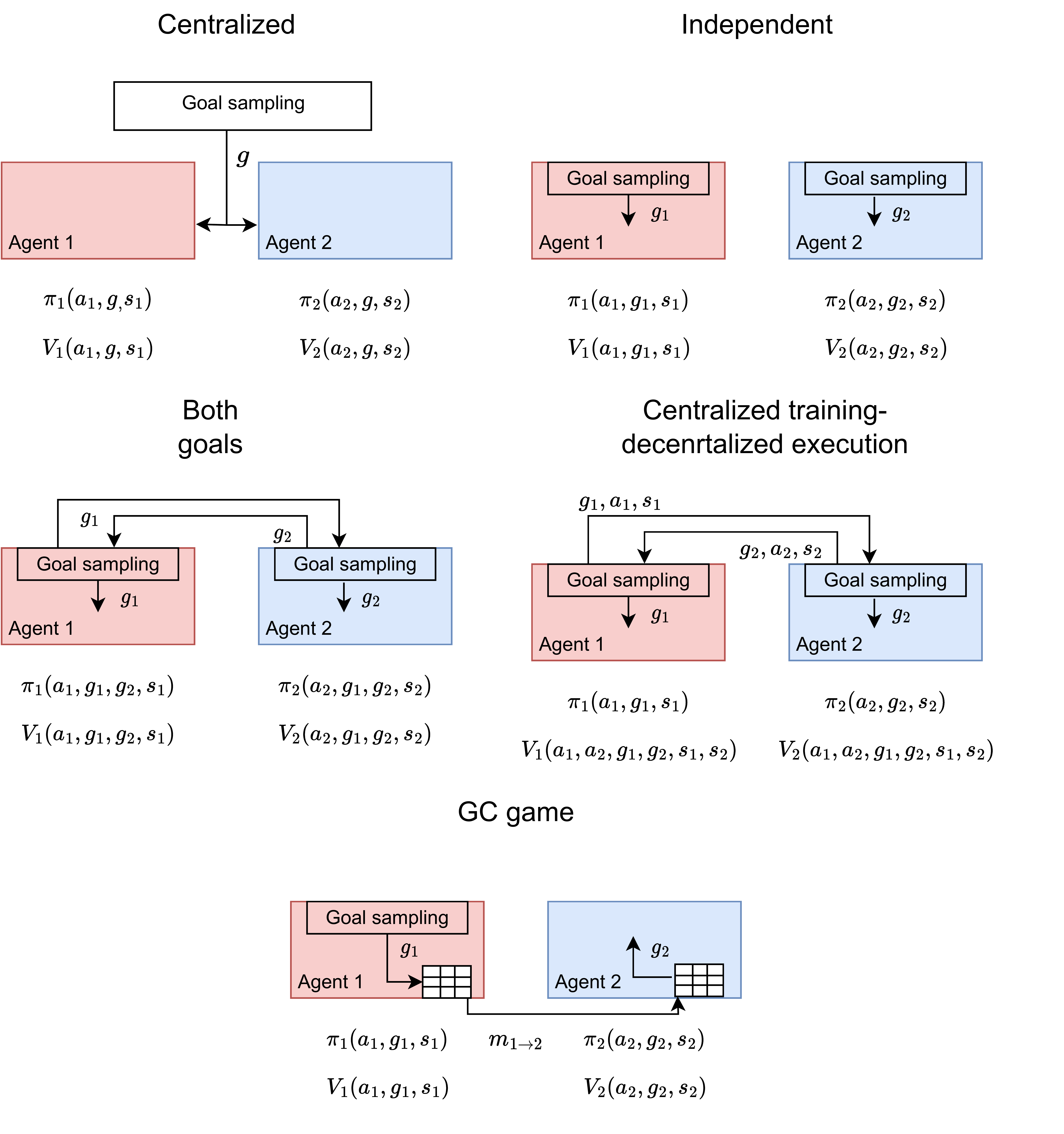}
\caption{Illustration of the different methods empirically evaluated in our work, where we indicate policies with $\pi$ and value functions with $V$ using the agent indexes as subscripts. The centralized baseline follows the algorithm proposed by \cite{yang2018cm3}, while the independent and both-goals baselines can be viewed as the equivalent of independent and joint learners proposed in the past~\citep{claus1998dynamics} but for goals instead of state-actions.}
\label{fig:baselines}
\end{figure*}

In Figure \ref{fig:baselines} we present an illustration of how the different methods we evaluate vary in terms of the information available to each agent and how it is used to condition its policy and value function.

{\color{revise} 
\section{Insights into training complexity}\label{app:training}

Understanding how our proposed solution performs as the number and difficulty of goals increases is useful for future applications of the \algname to more complex settings.
This algorithm is faced with the task of simultaneously learning how to align goals and learning how to solve them.
In contrast, the centralized baseline (100$\%$-aligned) is only faced with learning how to solve goals.
To disentangle the difficulty of these two tasks we here study the complexity of these two methods in terms of two parameters: a) the size of the goal space b) the difficulty of achieving goals. To disentangle these two effects we make the following comparisons: 

For a) we compare the training and evaluation performance of the two algorithms between the 3-landmarks and 6-landmarks environment when only cooperative goals are considered (thus only goals of equal difficulty).  For b) we compare the training and evaluation performance between a setting where we train on both individual and cooperative goals and a setting with only cooperative goals, both in the 6-landmarks environment. Since the total number of goals is 21 and there are 6 individual goals, the latter setting contains about 70$\%$ more difficult goals.

We present results in Figure \ref{fig:nlandmarks} for the effect of goal space size and in Figure \ref{fig:complexity} for the effect of goal difficulty. Regarding the goal space size, we observe that doubling the size of the goal space leads to about four times slower convergence. This is intuitive as the number of goals changes from 6 (in the 3-landmarks) to 21 (6-landmarks), so more time is required to master the goals. This is also true for the centralized baseline, meaning that the increase in complexity is due to the need to learn more policies, rather than the need to align more goals. Regarding the goal difficulty, we see that the algorithm learns to solve quicker the task that has both individual and cooperative goals. Thus, although the number of goals increased the convergence time decreased. This is because the individual goals are solved more easily and then facilitate solving the cooperative goals. The same behavior is observed for the centralized baseline.

\begin{figure}
    \centering
    \includegraphics[width=0.7\columnwidth]{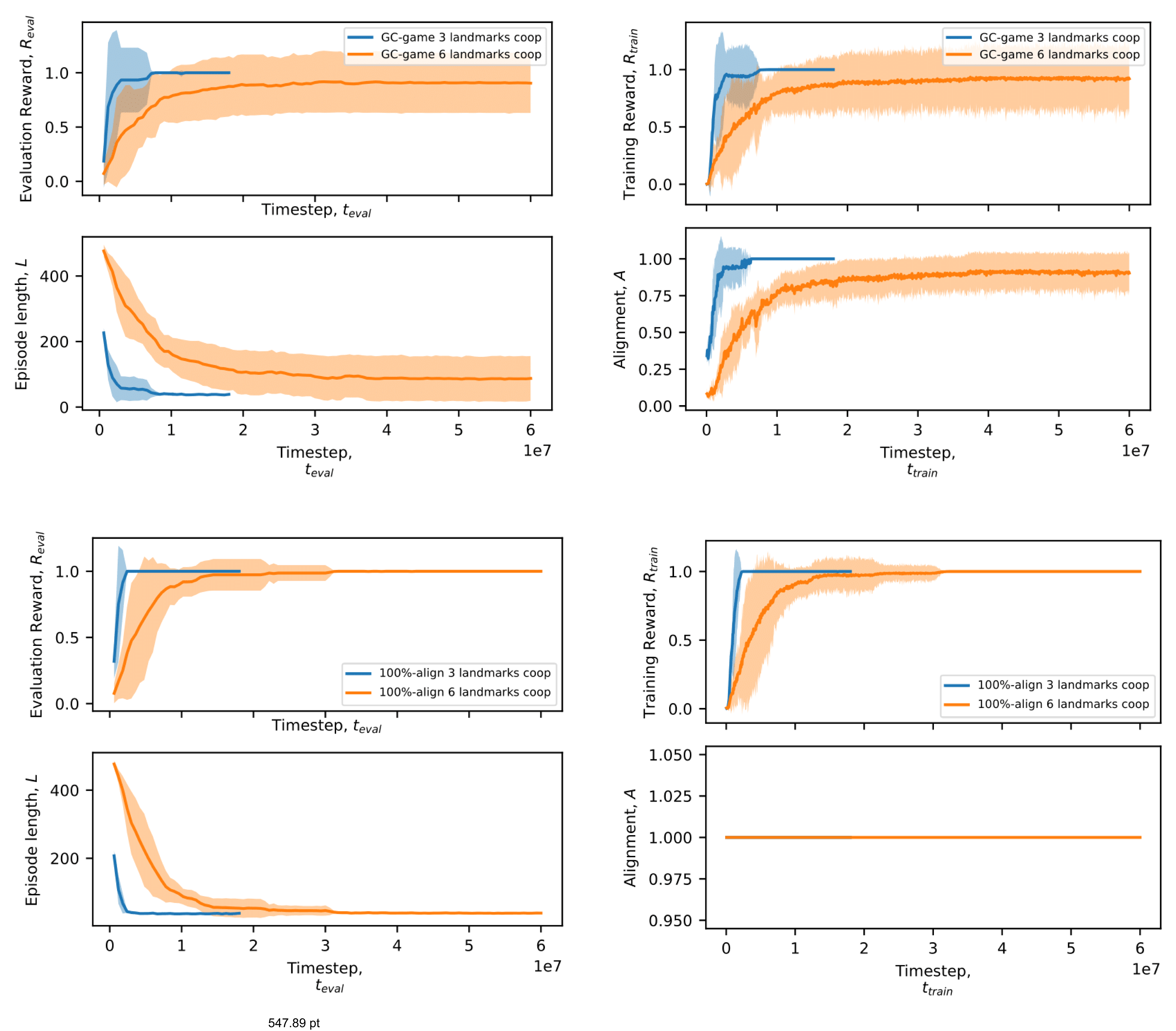}
    \caption{Computational complexity of the \algname and 100$\%$-aligned for environments with different numbers of landmarks}
    \label{fig:nlandmarks}
\end{figure}

\begin{figure}
    \centering
    \includegraphics[width=0.7\columnwidth]{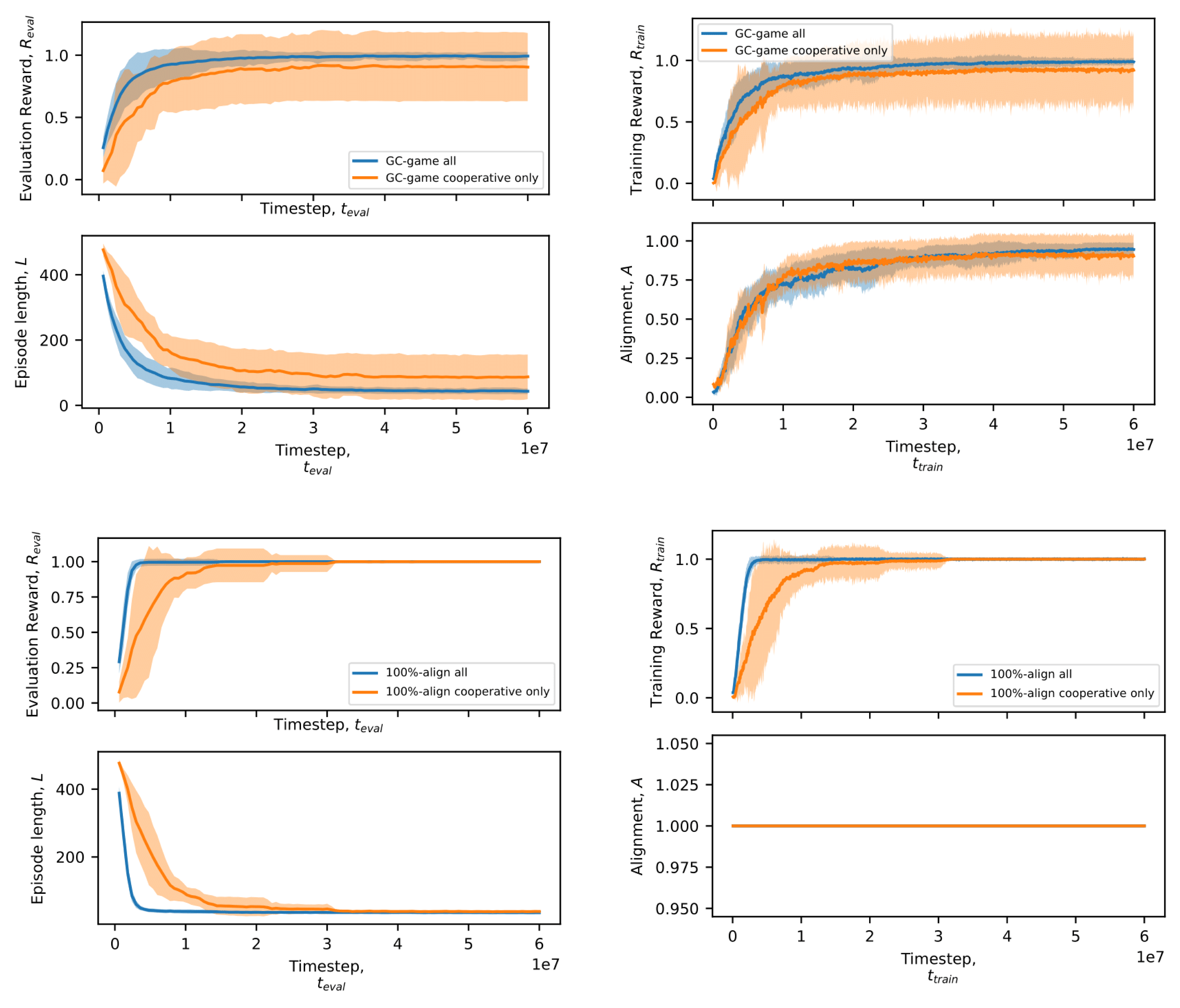}
    \caption{Computational complexity of the \algname and 100$\%$-aligned for environments with goals of different difficulty.}
    \label{fig:complexity}
\end{figure}

}

\section{Additional results}\label{app:results}

\subsection{3-landmarks environment}\label{app:3_landmarks}

Figure \ref{fig:3_landmarks} contains the evaluation performance, on the left, and training performance, on the right for the 3-landmarks environment. We observe that, compared to the 6-landmarks environment, the population requires significantly less training time (about one order of magnitude smaller) and that differences across methods during evaluation are not as pronounced. During training, we observe that alignment is correlated with performance with the independent baselines collecting the least rewards. Thus, we conclude that  our empirical conclusions generalize to simpler problem settings and that studying problems with increased task complexity is important for evaluating methods on the \probname.

\subsection{Effect of message size}\label{app:messages}

\begin{figure*}
\begin{minipage}{0.45\textwidth}
        \centering
    \includegraphics[width=0.7\columnwidth]{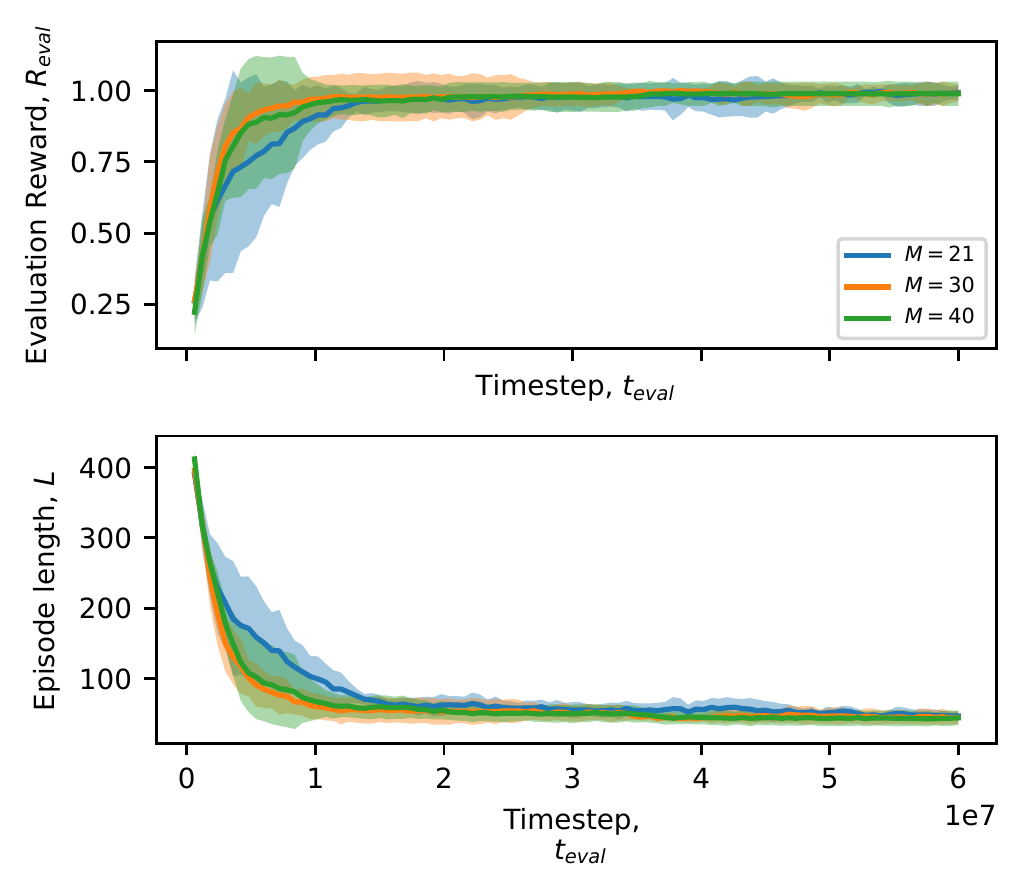}
\end{minipage}
\begin{minipage}{0.45\textwidth}
    \centering
\includegraphics[width=0.7\columnwidth]{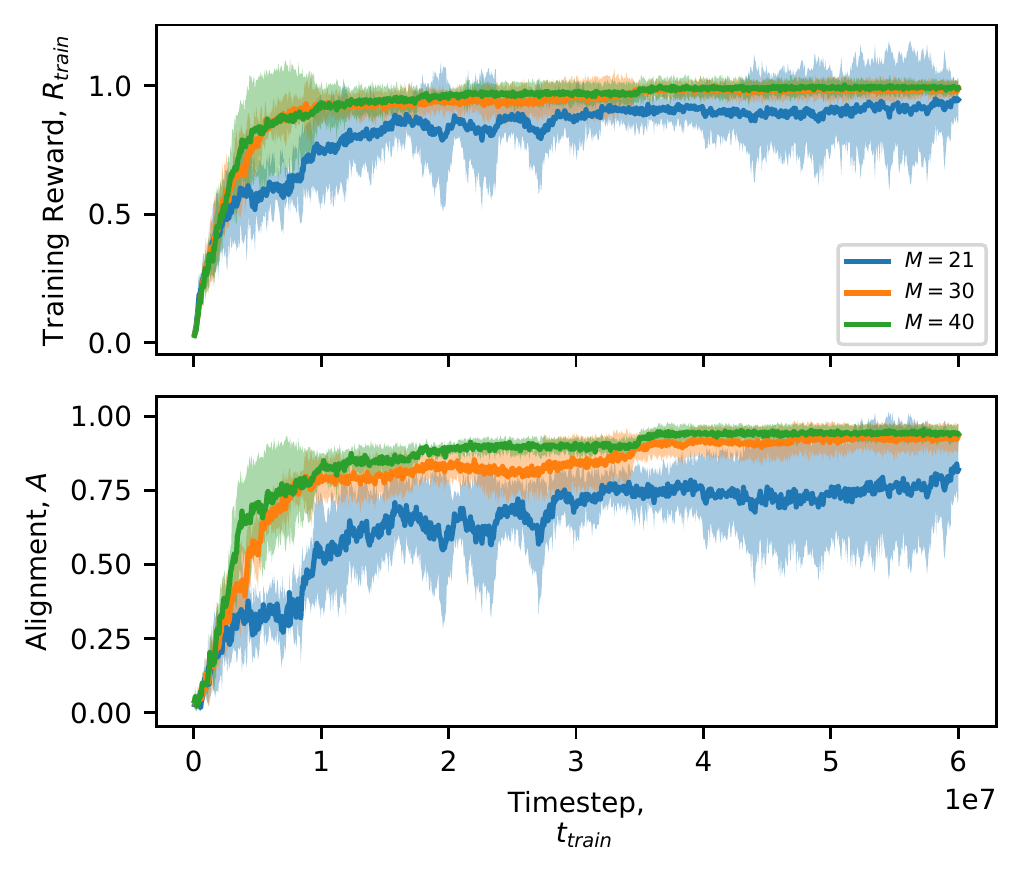}
\end{minipage}
\caption{Effect of message size $M$ on the \algname in the 6-landmarks environment during evaluation (left) and training (right) episodes }
\label{fig:messages}
\end{figure*}

{\color{revise}
In Figure \ref{fig:messages} we study the effect of message size  on the \algname in the 6-landmarks environment by setting it to the smallest possible value ($M=21$ is equal to the number of goals), a medium value ($M=30$) and a high value $(M=40)$. We observe that evaluation performance does not vary significantly with message size except for the fact that small message size leads to slower convergence to the optimal policy. During training, we observe that small message size cannot reach perfect alignment and amasses slightly lower rewards. Thus, we conclude that the message size should be set to a value relatively higher than the number of goals but no further benefits are gained when it increases beyond that.

The main observation here is that, during training, using a message size equal to the number of goals (21) leads to sub-optimal alignment and rewards. By observing the matrix tables for this specific example, we understood that this is due to a deadlock: if one message-goal association is learned incorrectly early in training (where incorrectly means that the goal of the leader and follower are misaligned) then a column/row is reserved and cannot be used for the correct association. This leads to at least two goals being misaligned until the end of training. When we slightly increase the number of messages, on the other hand, a wrong association can be fixed later in training, because there are still enough degrees of freedom to align all messages.}

\subsection{Effect of scaling factor $\beta$}\label{app:multiplier}

\begin{figure*}
\begin{minipage}{0.45\textwidth}
    \centering
    \includegraphics[width=0.7\columnwidth]{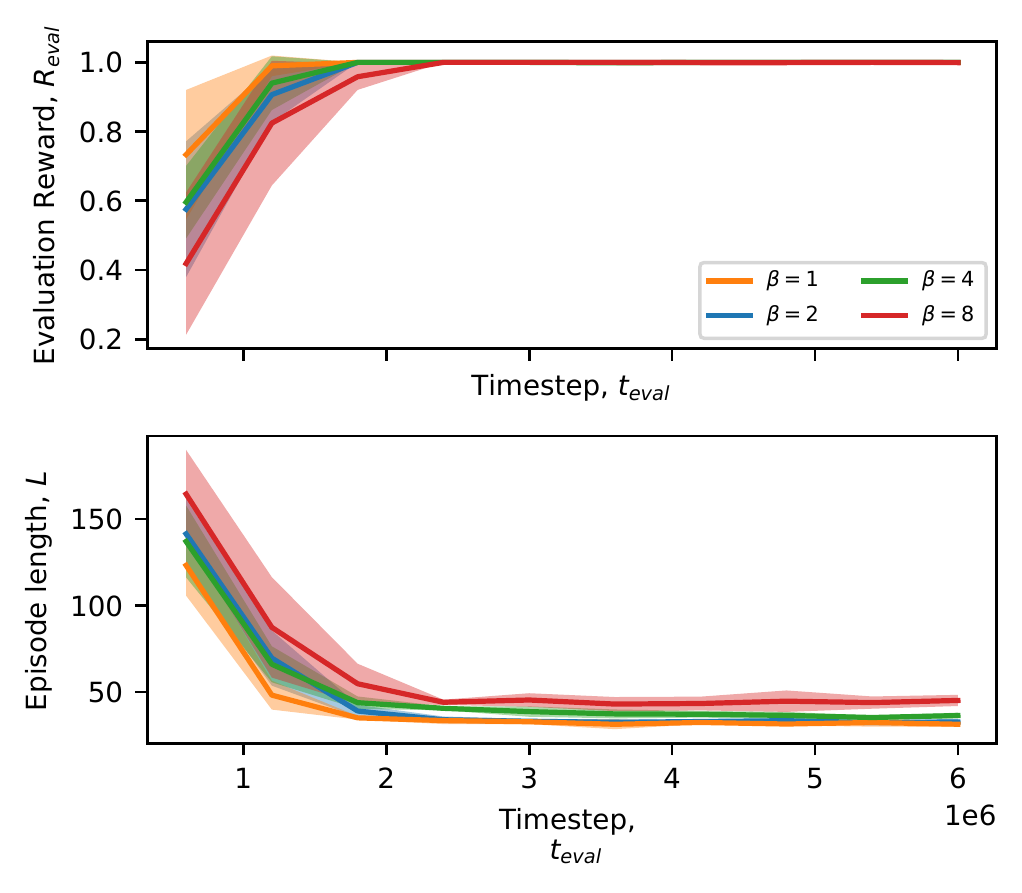}
\end{minipage}
\begin{minipage}{0.45\textwidth}
    \centering
    \includegraphics[width=0.7\columnwidth]{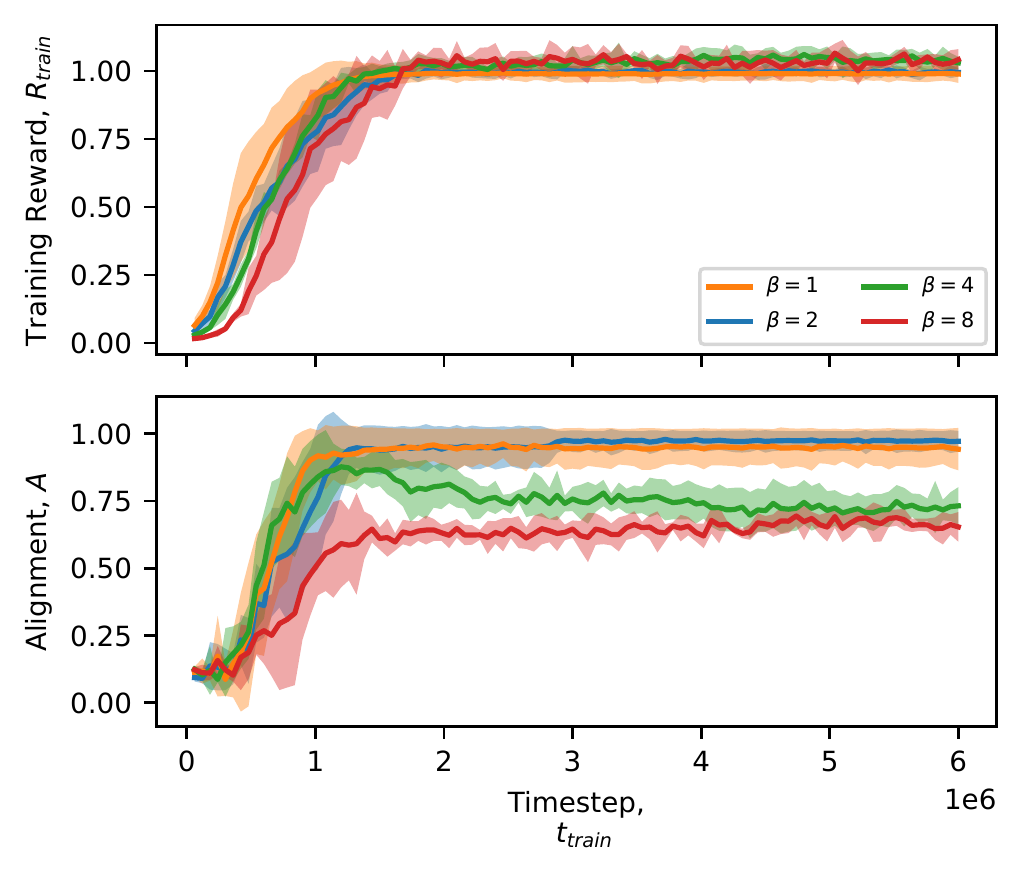}
\end{minipage}
\caption{Effect of scaling factor $\beta$ on the \algname in the 3-landmarks environment during evaluation (left) and training (right) episodes }
\label{fig:beta}
\end{figure*}

As we described in Section \ref{sec:results} the scaling factor $\beta$ controls the relative importance of individual versus cooperative goals: increasing the value of $\beta$ indicates a proportional decrease in the importance of solving independent goals. To examine the effect of $\beta$ we present the perfornance of the \algname for different values ($\beta \in [1,2,4,8]$) in Figure \ref{fig:beta}.
We observe that, for the \algname, higher values of $\beta$ lead to lower alignment: as cooperative goals are very rewarding in this case agents with the role of follower prefer them over individual ones even when the leader communicates about a cooperative goal.
At the same time, low values of $\beta$ lead to slower convergence to the optimal solution, as agents with the role of follower are not incentivized enough to choose cooperative goals, as they still receive rewards when they choose individual goals regardless of the leader's follower.
Finally, contrasting the behavior of the \algname to the other baselines in Figure \ref{fig:beta} shows that, by increasing $\beta$, the \algname can amass more rewards during training that the centralized baseline. This is not surprising: as the agents learn this risky behavior or aligning cooperative with individual goals, they experience more rewarding episodes.

{\color{revise}

\subsection{Recurrent policies}\label{app:recurrent}
We have so far employed only feedforward policies in all our methods. 
We now study the effect of using a recurrent policy.
Our intuition is that a recurrent policy can facilitate adaptation during the episode, as an agent can infer the direction the other is moving to and, perhaps, its goal. 
In Figure \ref{fig:recurrent} we compare this recurrent baseline with the other methods during evaluation and training trials.
We observe that, during training, the recurrent policy with independent sampling (Recurrent 0\% align) performs as badly as the independent feedforward baseline, while during evaluation, it is the worst-performing method.
Thus, introducing a recurrent policy did not facilitate adaptation.
Moreover, as the recurrent policy with centralized training (recurrent 100\% align) converged to maximum reward, we can conclude that even with a recurrent policy the noisy training signal impacts learning when sampling goals independently. Since this method cannot lead to alignment, it is also negatively impacted by the large number of infeasible episodes.
Finally, the fact that, with independent sampling, the recurrent policy performed worse than the feedforward one may mean that it may be even more sensitive to this noisy training signal.  
It is, however, possible that it may benefit from further hyper-parameter tuning, for example an increase in the size of the neural network.


\begin{figure*}
\begin{minipage}{0.45\textwidth}
    \centering
    \includegraphics[width=0.7\columnwidth]{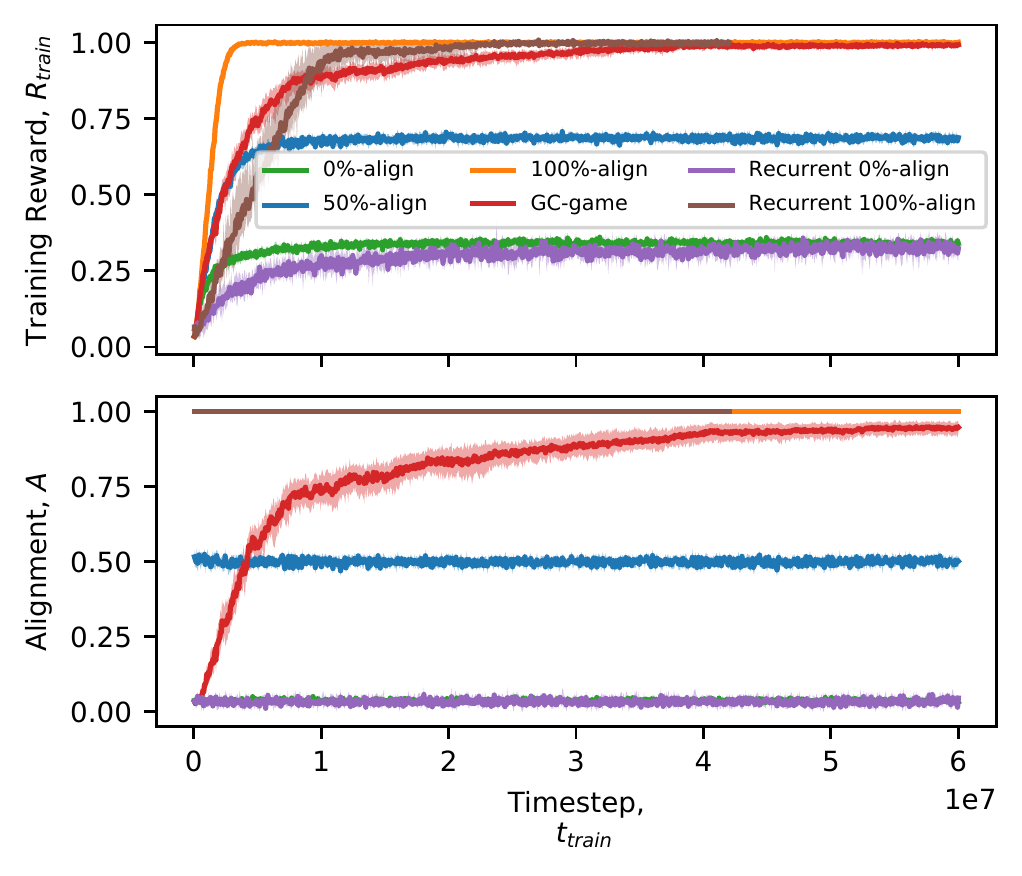}
\end{minipage}
\begin{minipage}{0.45\textwidth}
    \centering
    \includegraphics[width=0.7\columnwidth]{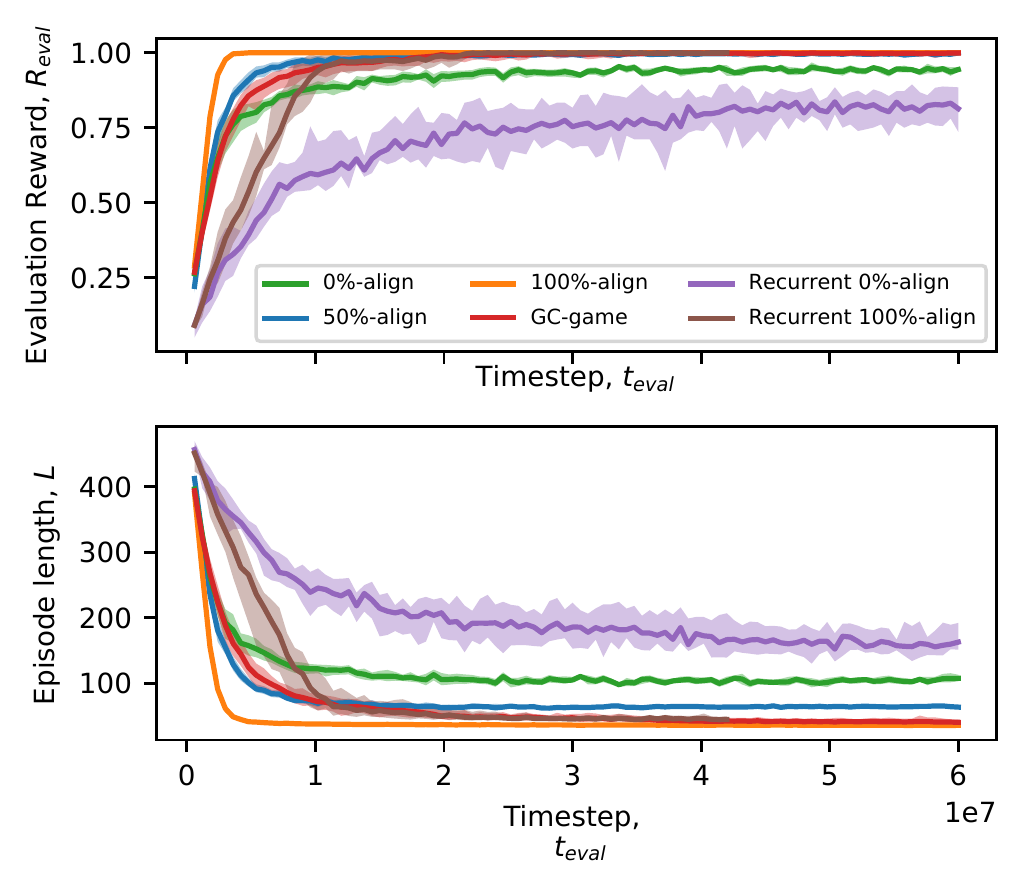}
\end{minipage}
    \caption{Comparison of the recurrent policy to the other baselines}
    \label{fig:recurrent}
\end{figure*}

\subsection{Learning progress for sampling goals}\label{app:LP}

Throughout the manuscript we have considered that, when an agent sets its own goal, it does so by randomly sampling within the goal space. 
This is the simplest form of intrinsic motivation.
An interesting question is how our independent baseline would behave if the sampling of goals is performed in a more sophisticated way, for example based on the competence of an agent.
Learning progress is such a type of intrinsic motivation that has been previously employed in single-agent settings~\citep{https://doi.org/10.48550/arxiv.1810.06284,kovac2022GRIMGEPLearningProgress}.
Here, we extend learning progress to our two-agent setting.
Our main motivation for this small study is to test whether introducing learning progress will indirectly lead to goal alignment.
Our intuition is that, by helping the agent focus on easier tasks first and then tackle the more challenging ones, this approach may lead to a curriculum from independent (easy) to cooperative (difficult) goals and that this may facilitate alignment.

At the start of an episode, each agent has a vector $LP 
\in [-1,1]^K$. Each coordinate of this vector is an approximation of the derivative in time of the competence of that agent for solving each goal. Goals are selected using a $\epsilon - \text{greedy}$ strategy and a proportional probability matching using the absolute value of the LP. For each agent $i$, the probability of selecting goal $g_i$ is given by:

\begin{align*}
    p(g_i) = \epsilon \times \frac{1}{K} + (1-\epsilon) \times \frac{|LP^a_{g_i}|}{\sum_{j=1}^{K}|LP^a_{g_i}|}
\end{align*}

The use of the absolute value makes agents concentrate both in goals that are currently being learned or forgotten. In the original implementation, LP is computed during evaluation rounds which provide a better signal than training data. However, in our multi-agent context, goal selection should be decentralized. Therefore, the first change we make to the strategy is to get rid of the evaluation rounds, and only estimate the learning progress based on experience from training. As we also want to work in a fully-decentralized context, at goal selection time we won't assume that one agent can have privileged information from the other (e.g access to other agent's goal). Each agent keeps a competence vector $C[n] \in [0,1]^{K}$ whose entries $C[n]_{i}$ is the moving average of the rewards obtained when the agent selected goal $i$ for the $n$ time during training. This average is defined by a window length $w$, which is the number of episodes we want to include for computing it. Then, the learning progress at time $n$ is:

\begin{align*}
    LP[n]_{i} = C[n]_{i} - C[n-w]_{i}
\end{align*}

\begin{figure}[H]
    \centering
    \includegraphics[width=0.7\linewidth]{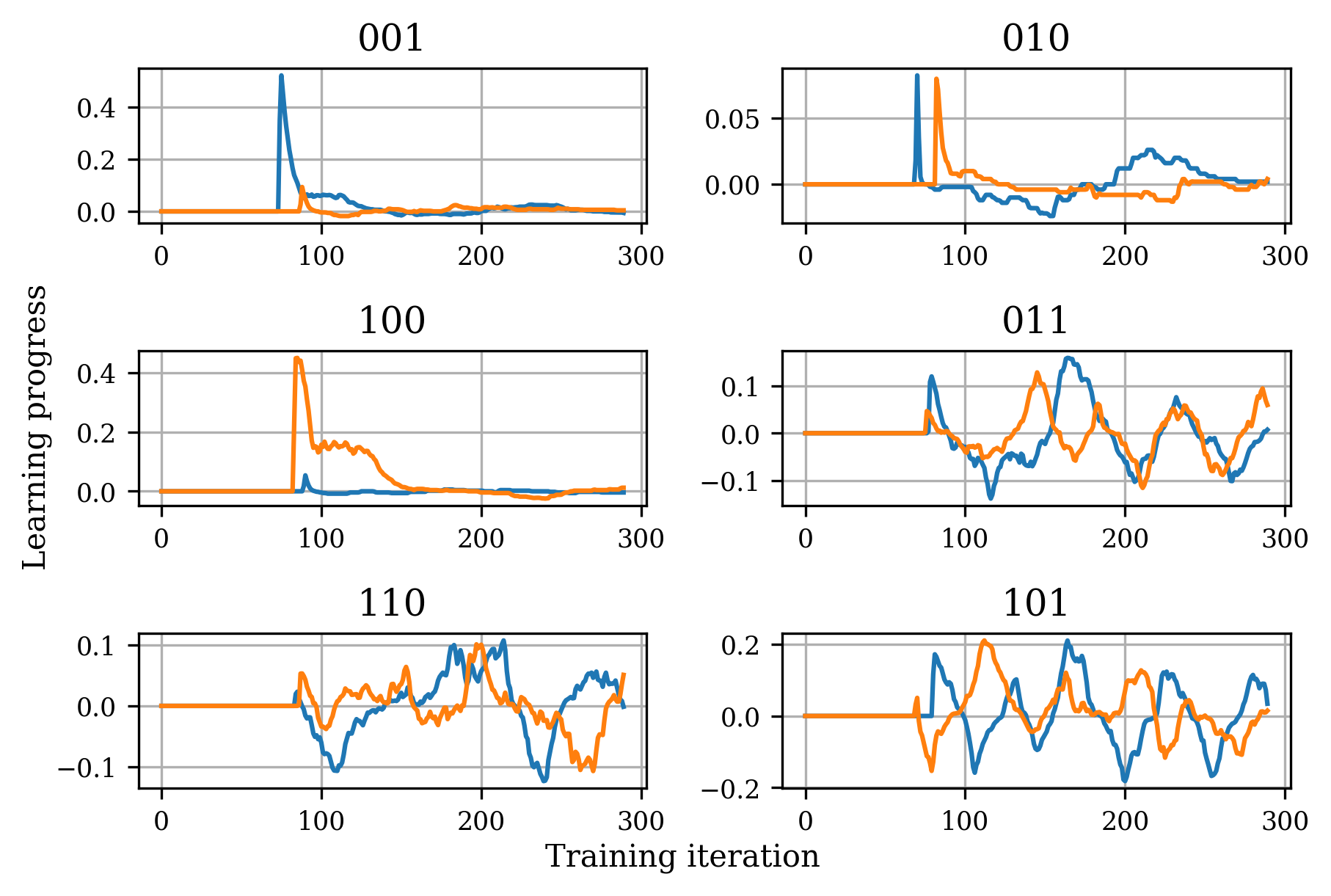}
    \caption{Learning progress estimate for each goal and two agents (blue and orange).}
    \label{fig:LP}
\end{figure}

Our experiments showed that estimating LP did not improve the performance of the independent baseline. As we observe in Figure \ref{fig:LP} LP values are very noisy and therefore lead to sampling goals relatively randomly, certainly not showing a curriculum from individual to cooperative goals. This is not surprising: estimating LP in multi-agent environments is challenging, because the competence of one agent depends on the competence of other agents as well. One agent's LP and competence in a cooperative goal doesn't only depend on the behavior of that agent, but also on the rest of them.
Furthermore, this estimate includes data from episodes where the pair of goals was impossible to solve (e.g one agent sampled one cooperative goal and the other an individual goal that cannot be solved at the same time). This is particularly evident for cooperative goals: for individual goals the LP plot look similar in shape to the ones presented in a previous work for a single-agent setting~\citep{https://doi.org/10.48550/arxiv.1810.06284}, while for cooperative goals the curves are too noisy and do not converge.
}

\subsection{Cooperative goals only}\label{app:coop_only}

\begin{figure*}
\begin{minipage}{0.45\textwidth}
    \centering
    \includegraphics[width=0.7\columnwidth]{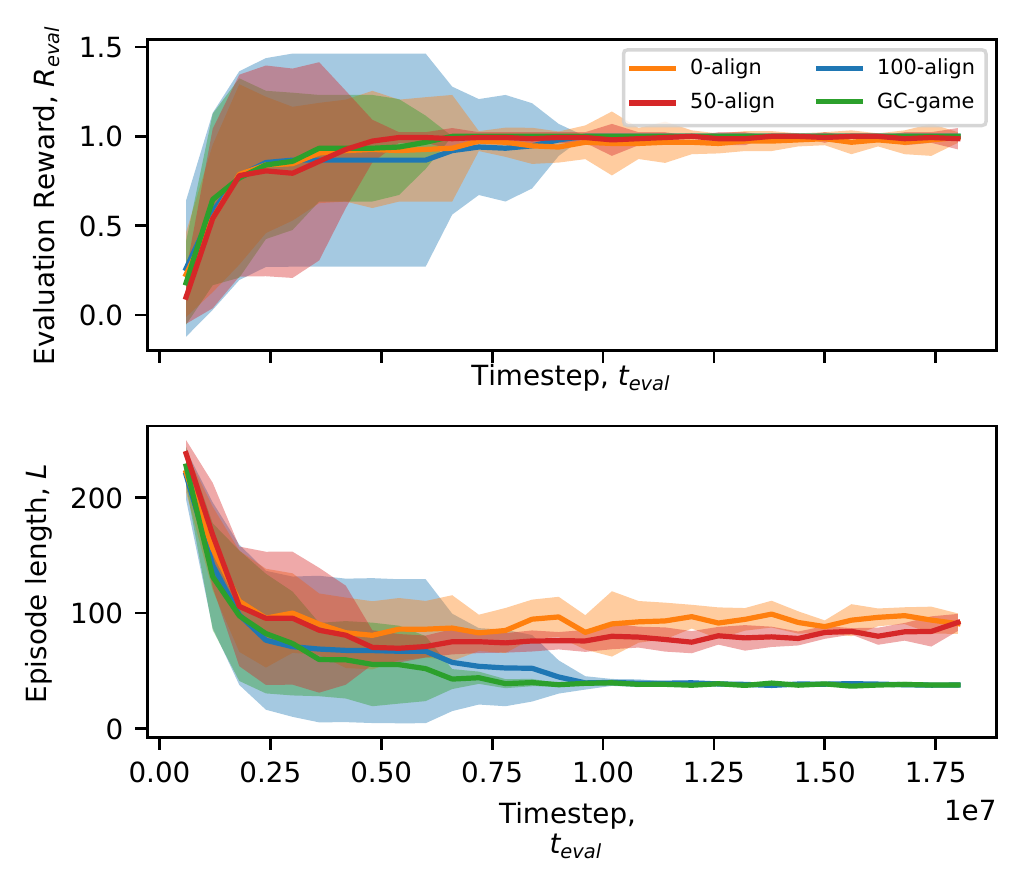}
\end{minipage}
\begin{minipage}{0.45\textwidth}
    \centering
    \includegraphics[width=0.7\columnwidth]{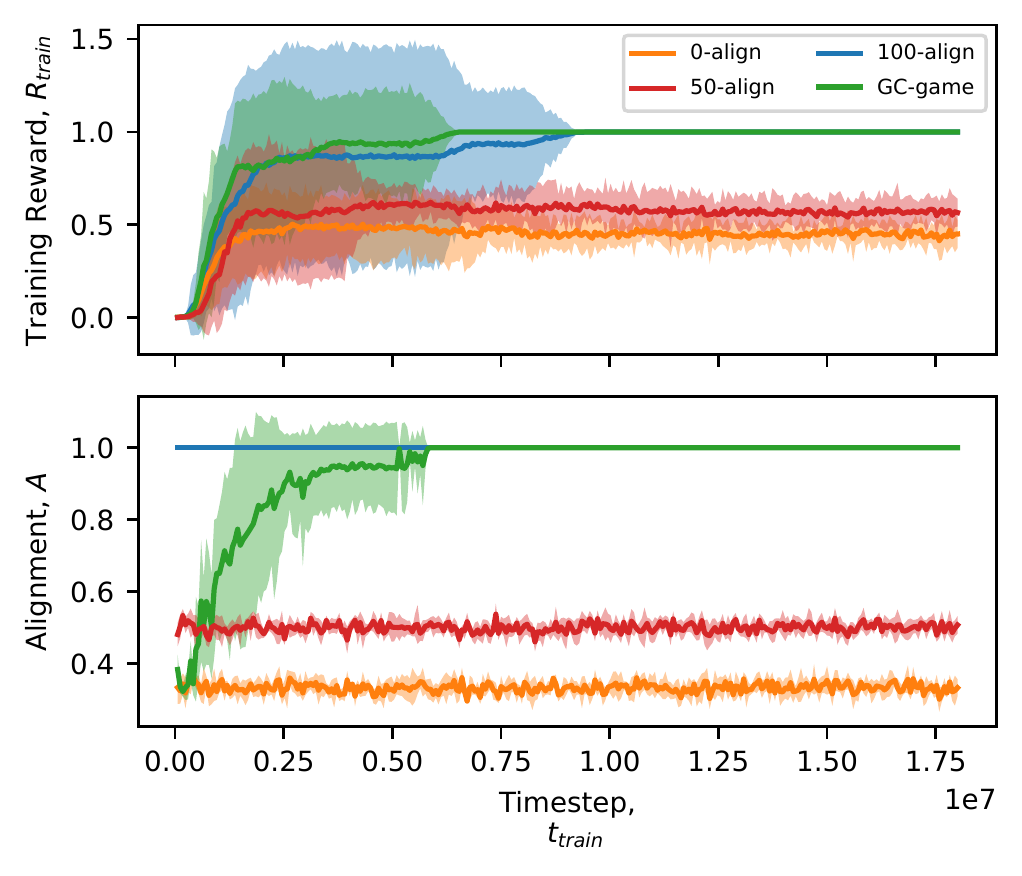}
\end{minipage}
\caption{Performance for the 3-landmarks environment with only cooperative goals during evaluation (left) and training (right) episodes for baselines
exhibiting different levels of alignment and the Goal-coordination game}
\label{fig:cooponly_3}
\end{figure*}

\begin{figure*}
\begin{minipage}{0.45\textwidth}
    \centering
    \includegraphics[width=0.7\columnwidth]{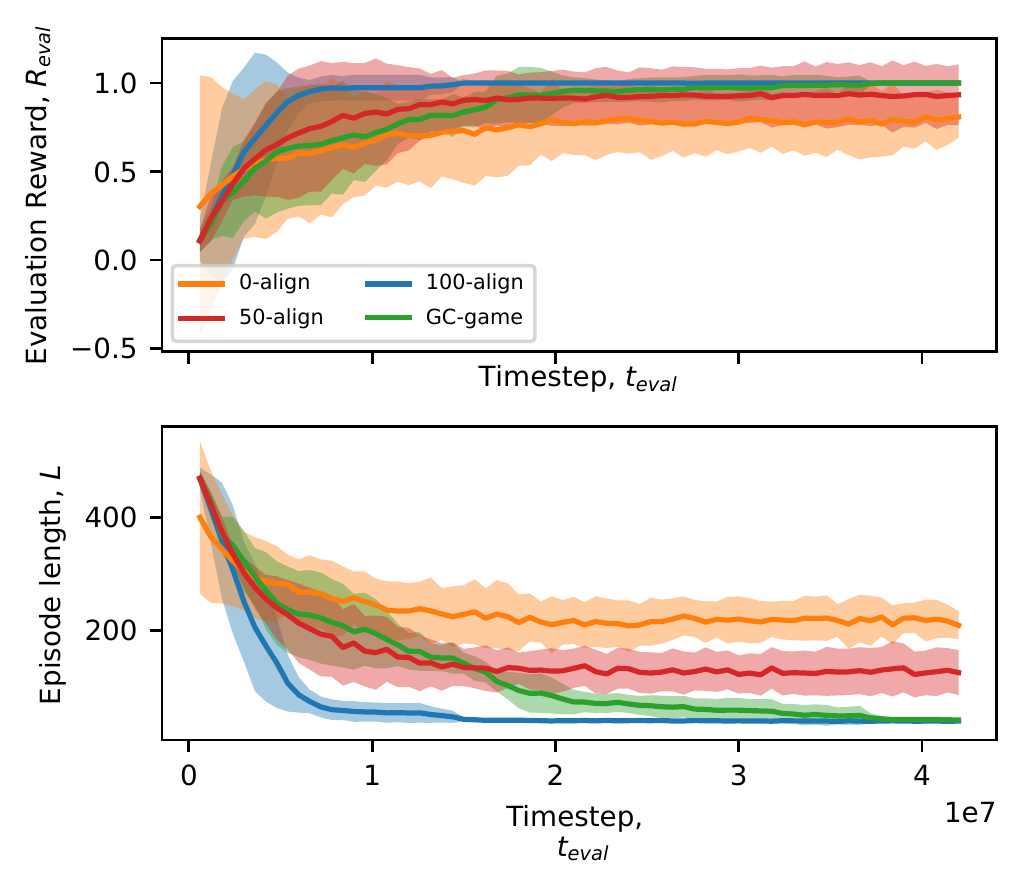}
\end{minipage}
\begin{minipage}{0.45\textwidth}
    \centering
    \includegraphics[width=0.7\columnwidth]{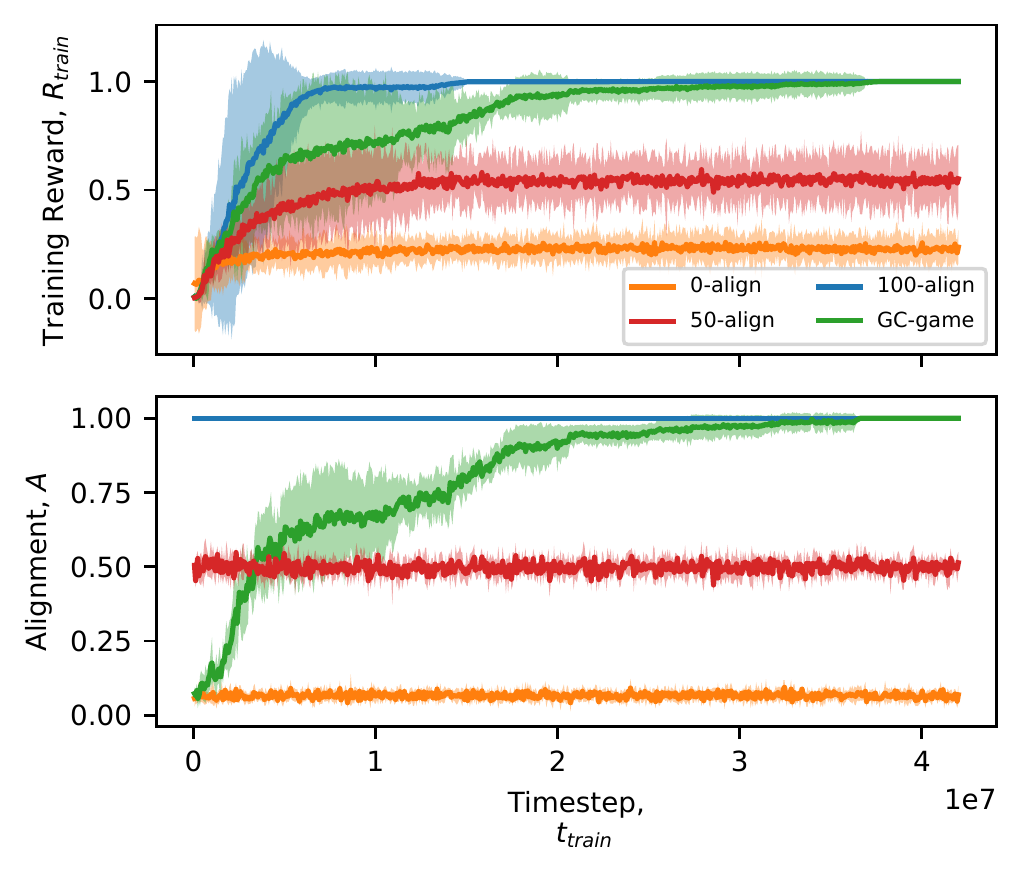}
\end{minipage}
\caption{Performance for the 6-landmarks environment with only cooperative goals during evaluation (left) and training (right) episodes for baselines
exhibiting different levels of alignment and the Goal-coordination game}
\label{fig:cooponly_6}
\end{figure*}

In this experiment, individual goals are removed both in training and evaluation. This means that the leader can only sample cooperative goals and the follower can only choose cooperative goals from its matrix. We observe the same conclusion as in the experiments with all goals but with bigger gap between methods both for the 3 landmarks \ref{fig:cooponly_3} and the 6 landmarks \ref{fig:cooponly_6} cases. Also in this setup we see that the \algname converges to $100\%$ alignment during training and converges to the same performances as the $100\%$ alignment method both in term of reward and episode length.

\subsection{Illustration of the "risky follower"}\label{app:risky}
\begin{figure}
    \centering
    \includegraphics[width=0.7\textwidth]{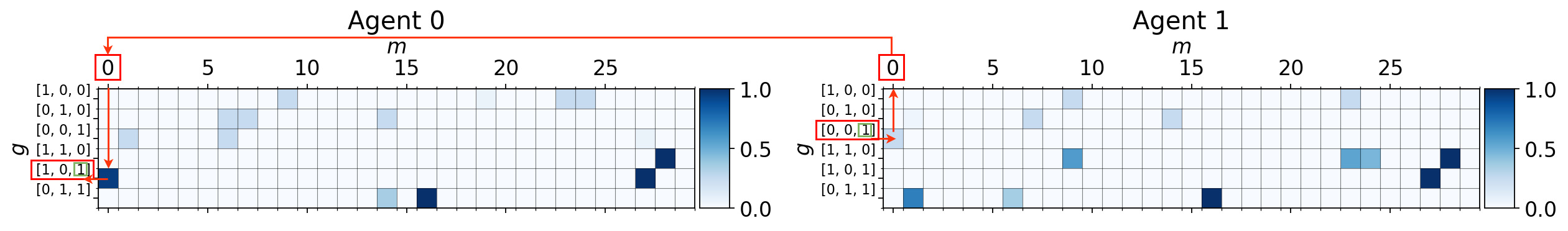}
    \caption{Example of the "Risky follower behavior" in the 3 landmarks case and reward multiplier $\beta =4$ . Leader is agent 1 which samples goal [0,0,1] and send message 0. Agent 1 is the follower and interprets message 0 as [1,0,1] which is cooperative and compatible with goal of agent 0.}
    \label{fig:risky_3land_matrix}
\end{figure}

We have described the "risky follower" behavior in Section \ref{sec:game_results}, where we defined it as a matching between a leader's individual goal and a follower's cooperative goal and presented the communication matrix that leads to it for the 6-landmarks environment. We now illustrate it for the 3-landmarks environment (with $\beta=4$) in Figure \ref{fig:risky_3land_matrix}. 

\begin{figure}
    \begin{minipage}{0.65\textwidth}
    \centering
    \includegraphics[width=0.7\textwidth]{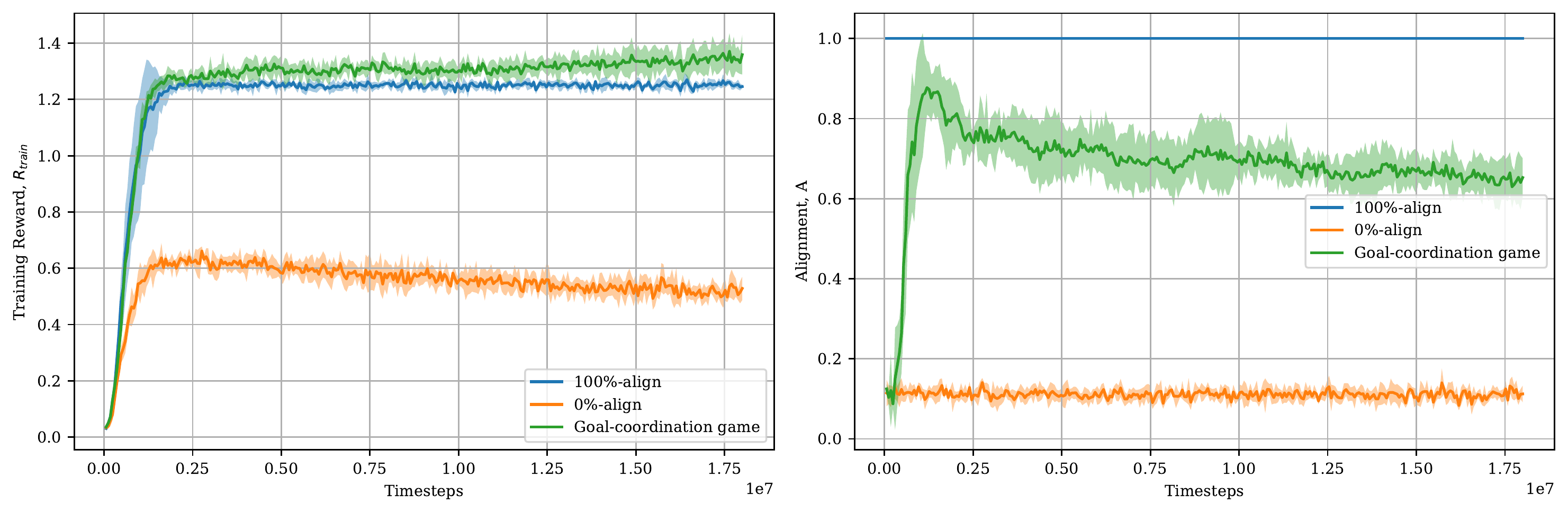}
    \caption{Training performances in the 3 landmarks case and $\beta=4$, we can see on left that goal coordination game exceeds the performances of the centralized (which attains its max value)}
    \label{fig:training_3landmarks_beta4}
    \end{minipage}\quad
    \begin{minipage}{0.32\textwidth}
    \centering
    \includegraphics[width=0.7\textwidth]{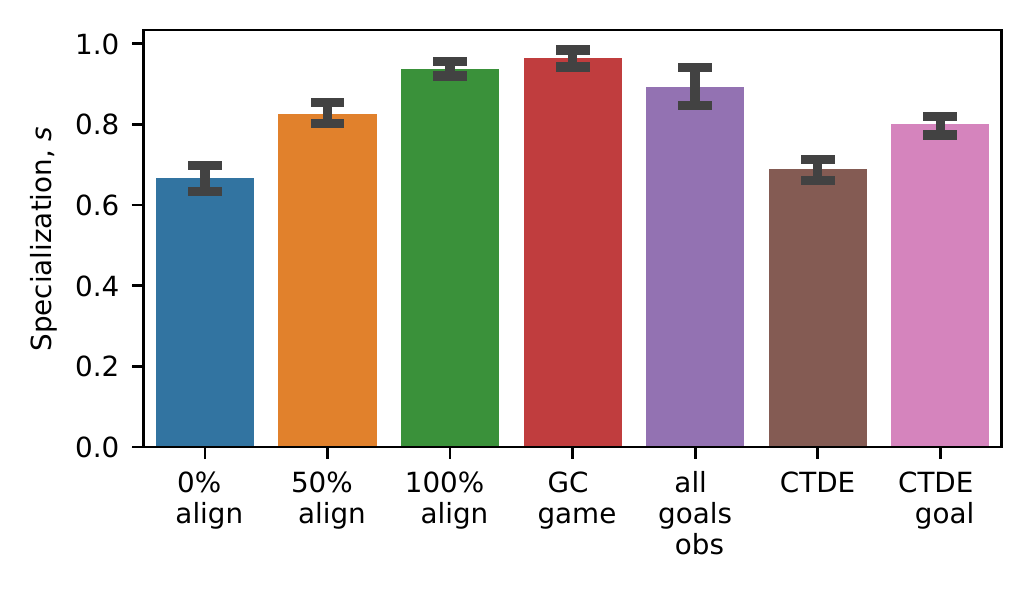}
    \caption{Specialization for the 6-landmark environment with $\beta=4$}
    \label{fig:special}    
\end{minipage} \
\end{figure}

We can even see on the training reward in Figure \ref{fig:training_3landmarks_beta4} that the risky follower behavior is used as we can see that the average reward of the goal coordination game is higher than the theoretical maximum of centralized training. In fact, in the case of the centralized training, the maximum average reward for one agent is capped by $P(sampling\_individual\_goal ) * R(individual\_goal\_fulfilled) +P(sampling\_cooperative\_goal)*R(cooperative\_goal\_fulfilled)= 0.5* 0.25 +0.5*1=0.625 $, and so the average reward of the sum of the 2 agents is capped by 1.25. While on the other hand, the \algname can allow the follower agent to get more reward when an individual goal is sampled by the leader and so have a higher maximum average sum of reward.


\subsection{Specialization}\label{app:special}
We defined specialization as the ratio of the episodes in which the agent went to its preferred landmark when following a cooperative goal in Section \ref{sec:role_align}. We now visualize the values of specialization we reported on the left of Figure \ref{fig:special}.

\begin{figure}

\begin{minipage}{0.45\textwidth}
    \centering
    \includegraphics[width=0.7\textwidth]{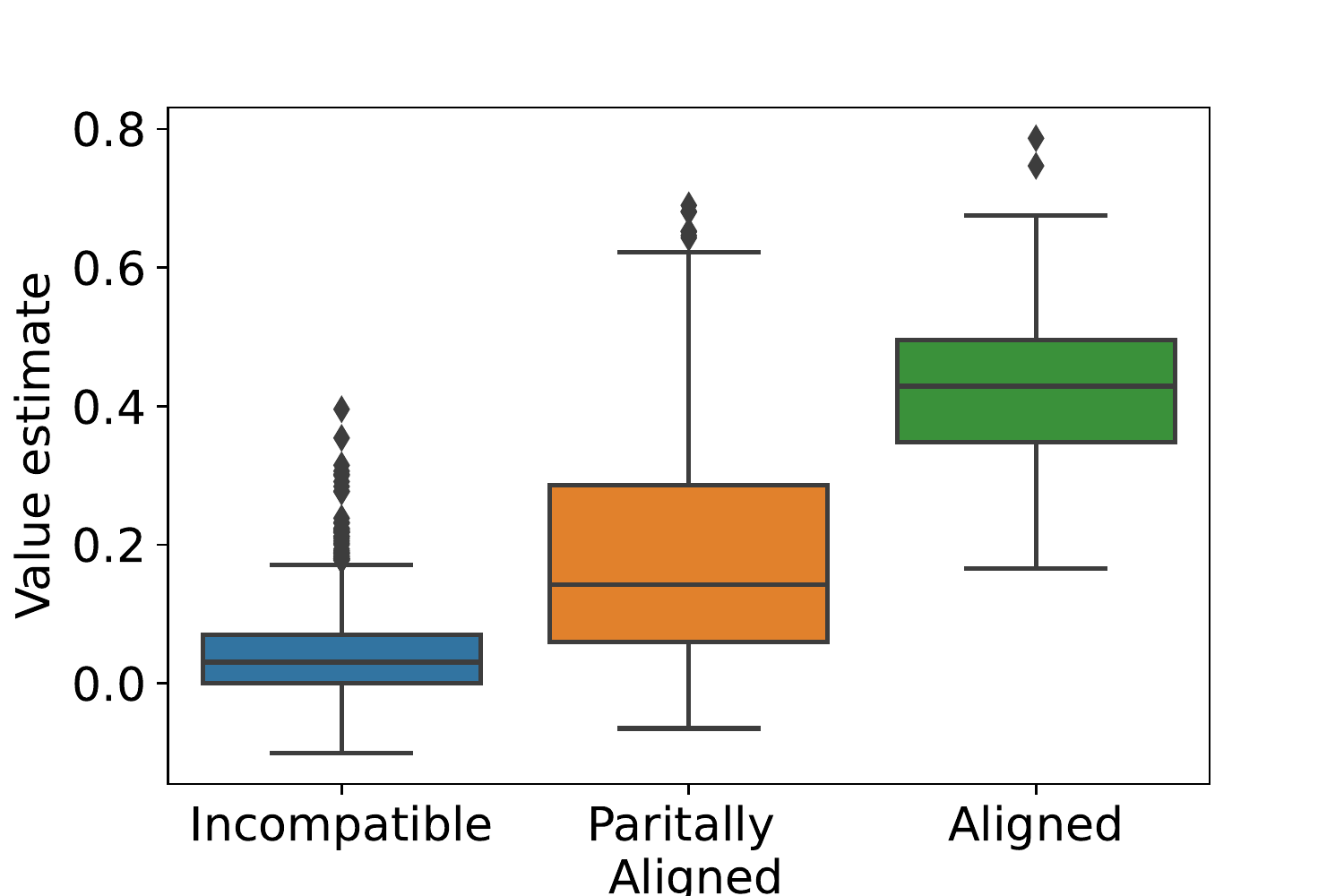}
    \caption{Learned value functions of the "both goal in the obs" baseline applied to different types of couple of goals. The boxplots distributions take into account several couple of goal and several seeds  }
    \label{fig:bothgoals_value}
\end{minipage} \quad
\begin{minipage}{0.45\textwidth}
    \centering
    \includegraphics[width=0.7\textwidth]{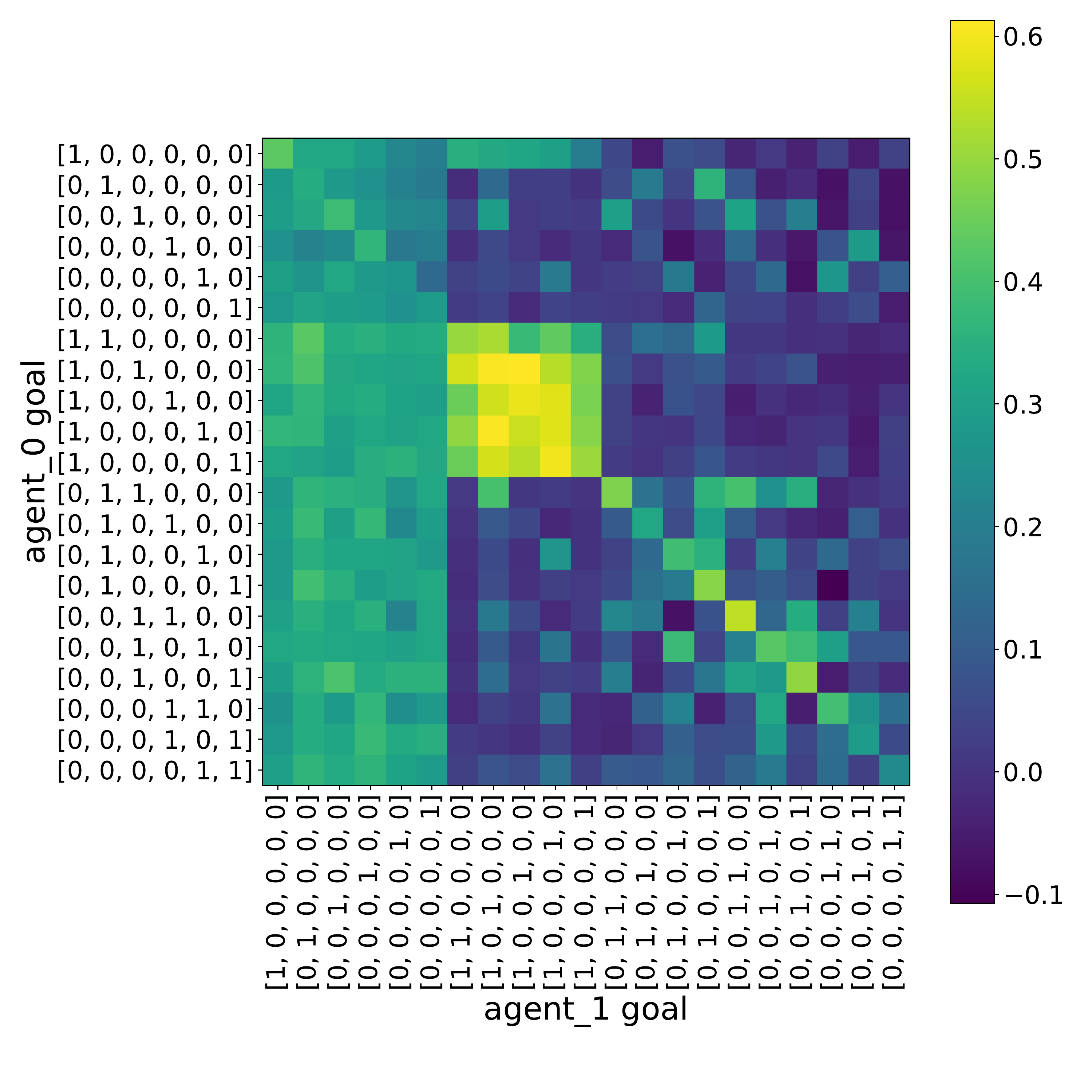}
    \caption{Value function of agent 1 (in one seed), in the both goals in the obs baseline, applied to different couples of goals for agent 0 and agent 1.}
    \label{fig:bothgoals_matrixvalue}   
\end{minipage}
\end{figure}



\begin{figure}
    \centering
    \includegraphics[width=0.7\textwidth]{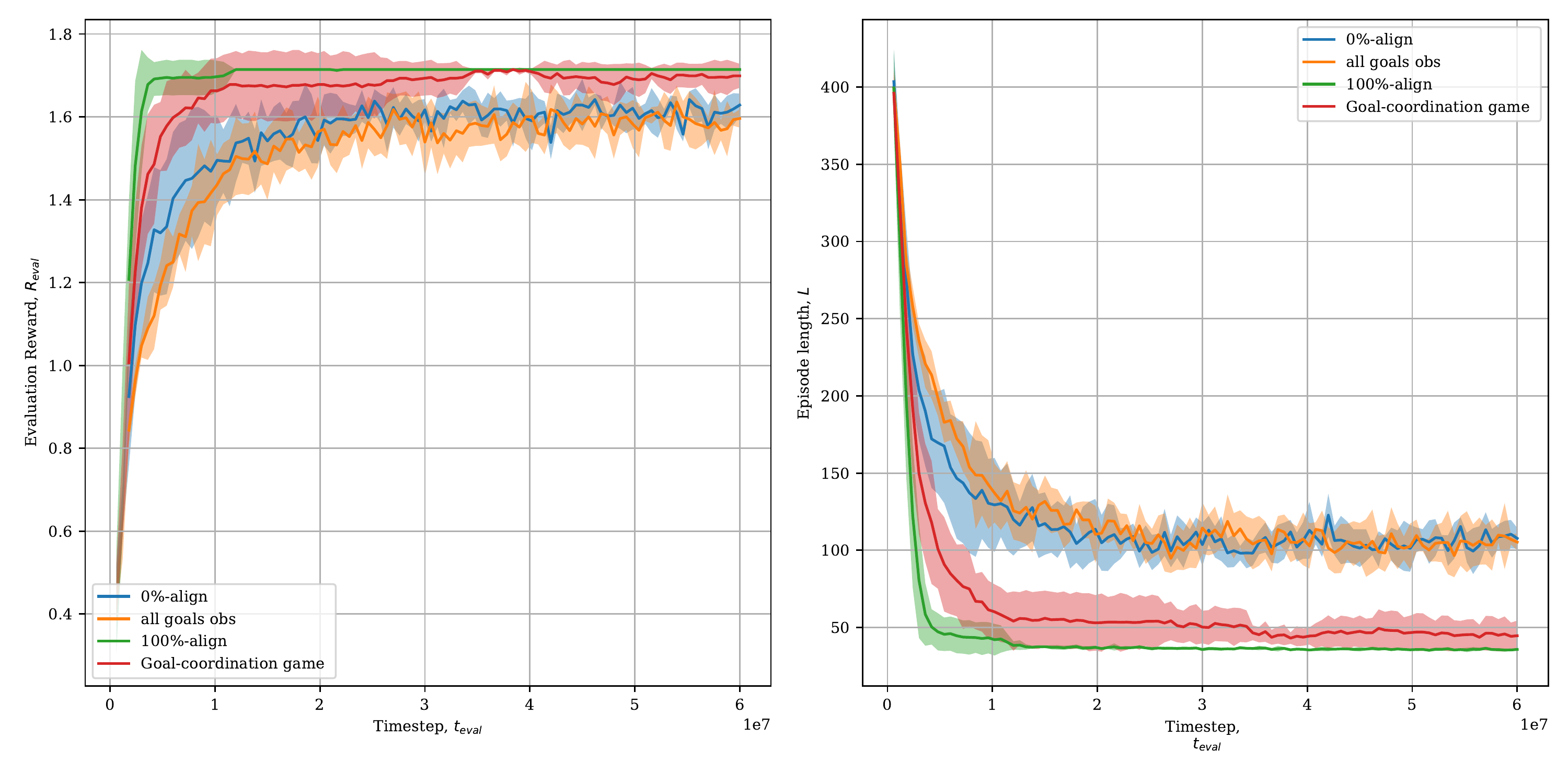}
    \caption{Comparing the performances of the all goals in the obs in evaluation, in the 6 landmarks, $\beta=2$ case}
    \label{fig:bothgoals_evalperf}    
\end{figure}

\subsection{Both goals observable}\label{app:bothgoals}

The objective of this experiment is to see if : a) agents that observe both goals can learn to ignore infeasible episodes b) agents that learn to ignore infeasible episodes  can achieve the same performance with the centralized baseline. This will help us understand if the independent baseline fails due to the noisy updates caused by infeasible episodes. As in our experiments the agent does not know the goal of the other agent, it might not know if the fail was due to it's policy or simply that the goal of the other was incompatible. By giving the goal of the other agents, we expect agents to learn to discard episode where goals are incompatible by giving them low value ( expecting no reward). 

In this section we thus study the case where every agent has access to both its goal and the goal of the other agent in the observation given to the policy and value network. We study this while being in the independent sampling case:  agents sample their goal independently of the other at the beginning of the episode.

In this section, we separate couples of cooperative goals into 3 categories: 1) incompatible goals are goals where there is no overlap at all, meaning that there is no common landmark in the cooperative goals of both agents (eg [1,1,0,0,0,0] and [0,0,1,1,0,0]; 2) Partially-aligned goals are cooperative goals which overlap on 1 of the landmark  (eg [1,1,0,0,0,0] and [0,1,1,0,0,0]); 3) aligned goals are when goals are the same. 

When looking at the distribution of value (given by the trained value function) on the different categories of couple of cooperative goals ( incompatible, partially aligned and aligned) across 5 seeds in Figure \ref{fig:bothgoals_value} , we can clearly see that the training learned to give very low value to incompatible goals and even to some of the partially compatible goals)

Even though training seems to learn that some goal associations are incompatible, and even though agents seem to specialize in Figure \ref{fig:special} the performances of the "Both goals observable" is still no better than independent without access to the goal of the other agent in Figure \ref{fig:bothgoals_evalperf}.

The fact that some values are high for the partially aligned goals case in Figure \ref{fig:bothgoals_value} is due to the specialization. If your goals overlap on only one landmark and you know that the other agent specialized to go to this common landmark when he has this goal, then you can go to your other landmark to get the reward. For example, looking at the matrix of agent 1 in one of the seed in Figure \ref{fig:bothgoals_matrixvalue}, we can see that when agent 0 has the goal [1,0,1,0,0,0], agent 1 has a high value for goals [1,1,0,0,0,0],[1,0,1,0,0,0],[1,0,0,1,0,0],[1,0,0,0,1,0],[1,0,0,0,0,1] which all contain landmark 1. This seems to indicate that agent 0 has a bias toward landmark 1 when having goal [1,0,1,0,0,0], which is exploited by agent 1. For example if agent 0 has goal [1,0,1,0,0,0] and agent 1 has goal [1,1,0,0,0,0], agent 1 will go to the 2nd landmark as he learned that agent 0 will go to the first one when he has goal [1,0,1,0,0,0].

\begin{figure}
\begin{minipage}{0.45\textwidth}
    \centering
    \includegraphics[width=0.7\textwidth]{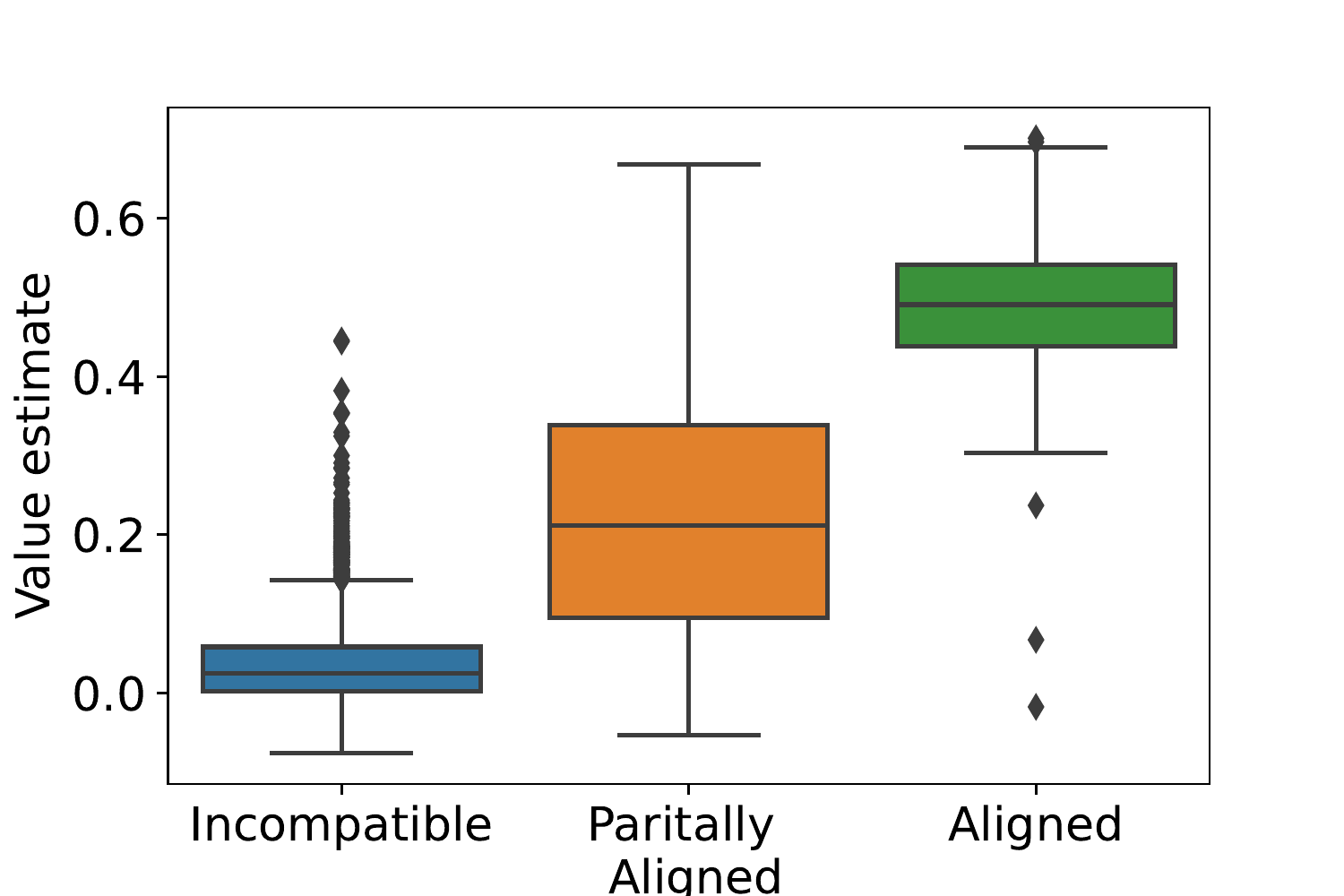}
    \caption{Learned value functions of the CTDE baseline applied to different types of couple of goals. The boxplots distributions take into account several couple of goal and several seeds }
    \label{fig:my_label}
\end{minipage} \quad
\begin{minipage}{0.45\textwidth}
    \centering
    \includegraphics[width=0.7\textwidth]{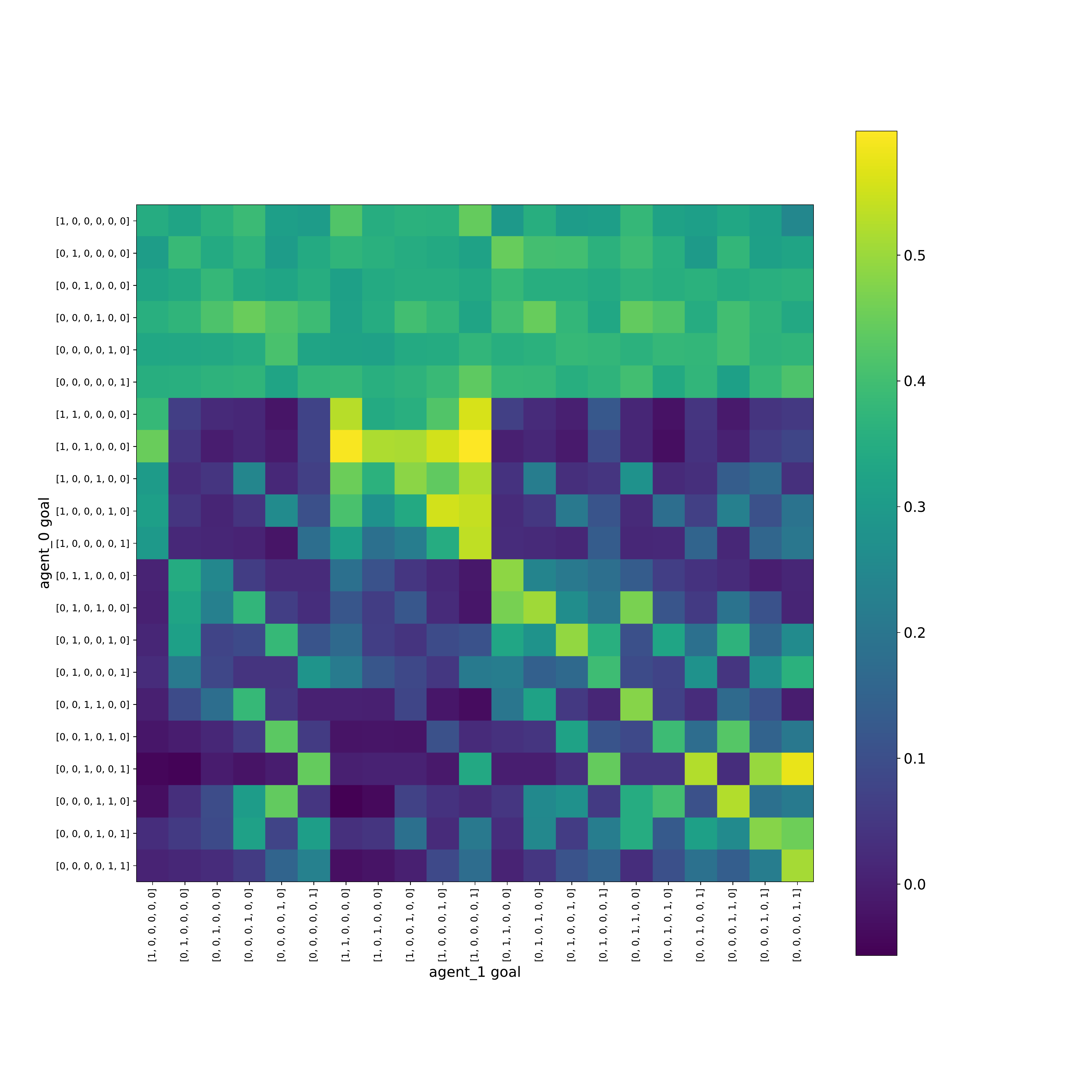}
    \caption{Value function of agent 0 (in one seed), in the CTDE baseline, applied to different couple of goals for agent 0 and agent 1}
    \label{fig:learned_infeasible}
\end{minipage}
\end{figure}

\begin{figure}[t]
    \centering
    \includegraphics[width=0.7\textwidth]{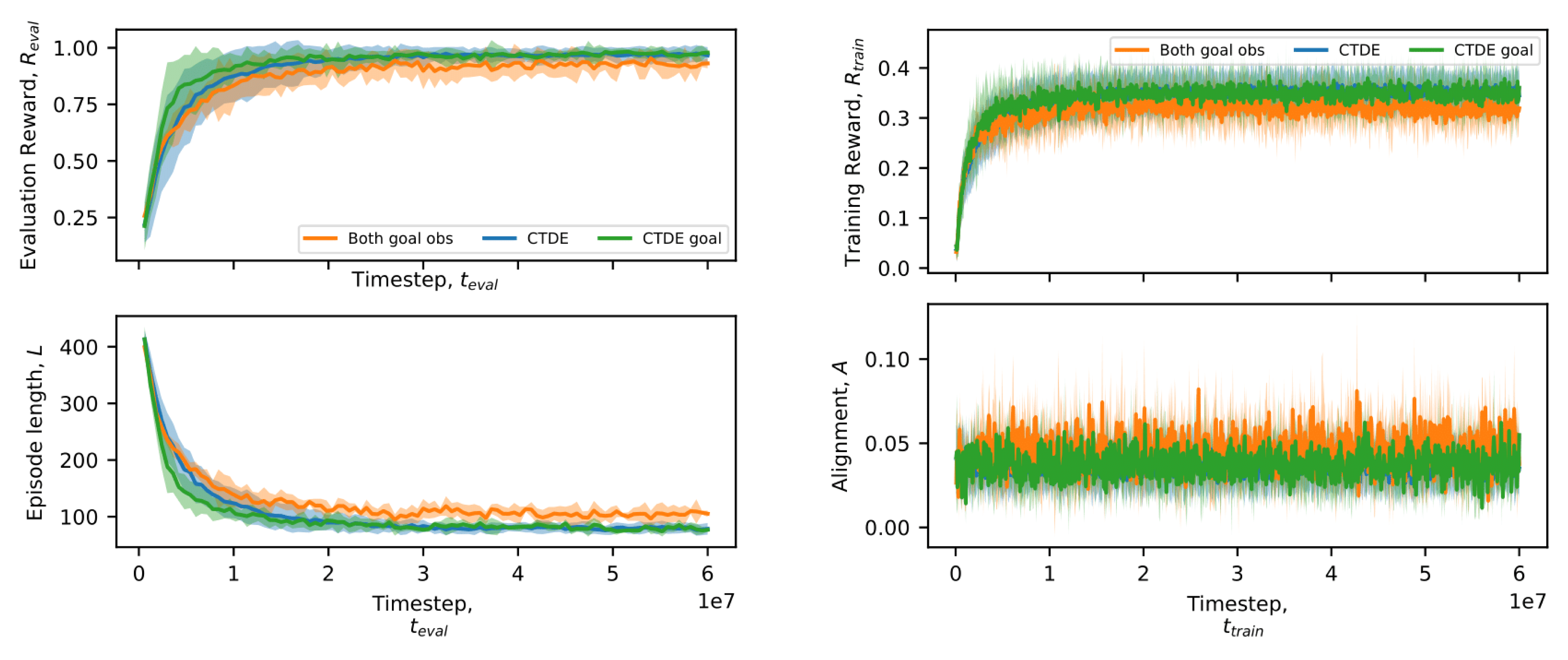}
    \caption{Comparing CTDE and two variants: the drop in performance happens when we condition the policies on both goals.}
    \label{fig:compare_CTDE}
\end{figure}

{ \color{revise} 
\subsection{Analysis of CTDE}\label{app:CTDE}
As we saw in the main paper in Section \ref{sec:results} in our discussion of Figure \ref{fig:performance}, the CTDE baseline performed better than the both-goals condition, which exhibited performance as bad as the independent baseline.
As we show in Figure \ref{fig:learned_infeasible}, CTDE has, similarly to the both-goals baseline learned how to detect infeasible episodes.
Why does this method outperform both-goals, then?
As we explained in our description of methods in Figure \ref{fig:baselines}, the two methods differ in two respects: a) the policies in both-goals are conditioned on both goals while only on an individual's goal for the CTDE b) the value function has access to, in addition to both goals, the states and actions of both agents.
To disentangle the effect of these two differences, we evaluate an additional method: a CTDE whose value function is conditioned on both goals but on an individual's state and action.
We compare the performance of these three slightly different methods in Figure \ref{fig:compare_CTDE}.
We observe that the new variant performs as well as CTDE, which indicates that the worse performance of both-goals is due to its conditioning on both goals.
This could be due to this method requiring further tuning, due to the increase in the size of the learning space, or due to the higher difficulty of learning policies conditioned on both goals.

To further understand the behavior of agents under CTDE, we also analyze the specialization of these variants and contrast it to the specialization of all baselines in Figure \ref{fig:special}, where we observe that CTDE has lower specialization.
In videos collected during evaluation trials, we also observed that CTDE agents exhibit intra-episode adaptation.
This indicates that CTDE is a promising approach towards unsupervised skill acquistion in our multi-agent settings.
Yet, it has not reached the performance of our proposed algorithm, although it introduces a need for centralization.
This is arguably due to the fact that this method is still plagued by the presence of infeasible episodes.


}

\end{document}